\documentclass[%
 aip,
 jcp,%
 amsmath,amssymb,
 preprint,%
]{revtex4-1}

\usepackage{amsmath}
\usepackage{graphicx}
\usepackage{dcolumn}
\usepackage{bm}

\draft 

\begin{document}


\title{Auger decay of molecular double core-hole state}


\author{Motomichi Tashiro}
\email[Author to whom correspondence should be addressed. Electronic mail:]{tashiro@ims.ac.jp}
\affiliation{Institute for Molecular Science, Nishigo-Naka 38, Myodaiji, Okazaki 444-8585, Japan}

\author{Kiyoshi Ueda}
\affiliation{Institute of Multidisciplinary Research for Advanced Materials, Tohoku University, Sendai 980-8577, Japan}

\author{Masahiro Ehara}
\affiliation{ Institute for Molecular Science, Nishigo-Naka 38, Myodaiji, Okazaki 444-8585, Japan}



\date{\today}

\begin{abstract}
We report on theoretical Auger electron kinetic energy distribution originated 
from sequential two-step Auger decays of molecular double core-hole (DCH) state, 
using CH$_4$, NH$_3$ and H$_2$CO molecules as representative examples. 
For CH$_4$ and NH$_3$ molecules, the DCH state has an empty 1s 
inner-shell orbital and its Auger spectrum has two well separated components.  
One is originated from the 1st Auger transition from the DCH state to the 
triply ionized states with one core hole and two valence holes (CVV states) 
and the other is originated from the 2nd Auger transition from the CVV states 
to quadruply valence ionized (VVVV) states. 
Our result on the NH$_3$ Auger spectrum is consistent with the experimental spectrum of 
the DCH Auger decay observed recently [Phys. Rev. Lett. {\bf 105}, 213005 (2010)].
In contrast to CH$_4$ and NH$_3$ molecules, H$_2$CO has four different DCH 
states with C1s$^{-2}$, O1s$^{-2}$ and C1s$^{-1}$O1s$^{-1}$ (singlet and triplet) 
configurations, and its Auger spectrum has more complicated structure compared to the 
Auger spectra of CH$_4$ and NH$_3$ molecules. 
In the H$_2$CO Auger spectra, the C1s$^{-1}$O1s$^{-1}$ DCH $\to$ CVV Auger spectrum  
and the CVV $\to$ VVVV Auger spectrum overlap each other, which suggests that 
isolation of these Auger components may be difficult in experiment. 
The C1s$^{-2}$ and O1s$^{-2}$ DCH $\to$ CVV Auger components  
are separated from the other components in the H$_2$CO Auger spectra, and can be observed in experiment. 
Two-dimensional Auger spectrum, representing a probability of finding two Auger electrons 
at specific pair of energies, may be obtained by four-electron coincidence detection 
technique in experiment. Our calculation shows that this two-dimensional spectrum is 
useful in understanding 
contributions of CVV and VVVV states to the Auger decay of molecular DCH states.  
\end{abstract}



\maketitle


\section{Introduction}


Molecular double core-hole (DCH) state is a state of a molecule with two inner-shell vacancies;  
they mostly correspond to two K-shell vacancies in the literature,\cite{ISI:A1986F086500030,ISI:A1987G075000061,ISI:A1993KZ44800006,ISI:000267697900020,ISI:000277756500014,ISI:000281296900043,Kryzhevoi2011,ISI:000281072100004,ISI:000281072100003,ISI:000284407400011,ISI:000287357100010}
although 
two L-shell vacancies or one K-shell and one L-shell vacancies have been treated as well.\cite{ISI:000291472400004} 
Two different kinds of molecular DCH state exist: single-site (ss) DCH state having 
core-hole vacancies at the same atomic site and two-site (ts) DCH state with two core-hole vacancies 
at different atomic sites. 
Properties of molecular DCH state with K-shell vacancies was 
first studied by Cederbaum et al.\cite{ISI:A1986F086500030} at 1986 and have been discussed 
occasionally since then. 
\cite{ISI:A1987G075000061,ISI:A1993KZ44800006,ISI:000267697900020,ISI:000277756500014,ISI:000281296900043,Kryzhevoi2011} 
In contrast, experimental realization of probing molecular DCH state has not been 
possible until recently because of technological difficulties. 
Recently, X-ray free electron laser (XFEL) at Linac Coherent Light Source (LCLS) started 
operation,\cite{ISI:000281467900020} which has capability to generate high intensity, 
short laser pulses required to produce molecular DCH state by sequential two-photon two-electron ionization. 
Using this XFEL facility, Fang et al.\cite{ISI:000281072100004} identified the 
ss-DCH state of nitrogen molecule by analyzing photoelectron and Auger electron spectra. 
Cryan et al.\cite{ISI:000281072100003} also obtained the Auger electron angular distribution 
originated from the N$_2$ DCH Auger decay in the molecular frame.  
%
At the same time, it has become possible to study molecular DCH state using 
synchrotron radiation (SR), combined with multi-electron coincidence technique. 
In contrast to two-photon sequential ionization by XFEL, 
SR mainly creates molecular ss-DCH state through single-photon two-electron ionization process. 
Eland et al.\cite{ISI:000284407400011} detected the ss-DCH states of 
NH$_3$ and CH$_4$ molecules in their SR experiment. For NH$_3$ molecule, they obtained the Auger electron 
spectrum originated from the cascade Auger decay of the ss-DCH state. 
The 2h-1p pre-edge resonance state of NH$_3$, with double 
core-hole single electron valence excited configuration, was also identified, 
where they found that the state decays predominantly through the spectator Auger process. 
In different SR experiment, Lablanquie et al.\cite{ISI:000287357100010} 
determined binding energies of the ss-DCH states for 
N$_2$, O$_2$, CO and CO$_2$ molecules. 
Using the four-electron coincidence detection method, with two photoelectrons and two Auger electrons, 
they obtained two-dimensional (2D) Auger intensity distribution of the N$_2$ ss-DCH decay as 
functions of two Auger electron kinetic energies.   
Their photoelectron spectrum of N$_2$ shows clear signature of 
the DCH shake-up satellite states. 
Based on the Auger spectrum of the N$_2$ DCH satellite state, they concluded 
that the satellite state decays by the spectator Auger process. 
%
%
%
%
Although number of experimental studies on molecular DCH state is still limited, 
it is expected to increase in near future because XFEL facilities are being 
constructed in Japan and EU,\cite{ISI:000284753200011} 
and also because more SR experiment will be performed to study molecular DCH state. 



Several theoretical studies 
\cite{ISI:A1986F086500030,ISI:A1987G075000061,ISI:A1993KZ44800006,ISI:000267697900020,ISI:000277756500014,ISI:000281296900043,ISI:000287482500008,Kryzhevoi2011} 
have been performed on molecular DCH state to clarify its basic properties. 
As in single core-hole state, molecular DCH state is not stable but decays by Auger 
electron emission or X-ray emission, where Auger decay is expected to be dominant for 
low-Z elements.\cite{ISI:A1991FV18400032} 
At present, number of theoretical works on Auger decay of molecular DCH state is quite limited, 
where only two studies exist: theoretical N$_2$ DCH Auger spectrum by Fang 
et al.\cite{ISI:000281072100004} 
and characterization of transient and final ion states in CH$_4$ and NH$_3$ DCH 
Auger decays by Eland et al.\cite{ISI:000284407400011} 
Detailed knowledge about Auger electron kinetic energy distribution, such as assignment 
of peaks and prediction of Auger intensity, is important to interpret experimental molecular 
DCH Auger spectrum, in which single core-hole Auger spectrum or other instrumental noise 
often prevents straitforward extraction of DCH Auger spectrum.\cite{ISI:000284407400011} 
In addition, detailed theoretical information may be valuable for 
studying time-dependent process of molecular DCH formation and decay in XFEL experiment.

\begin{figure}
 \includegraphics[scale=0.75]{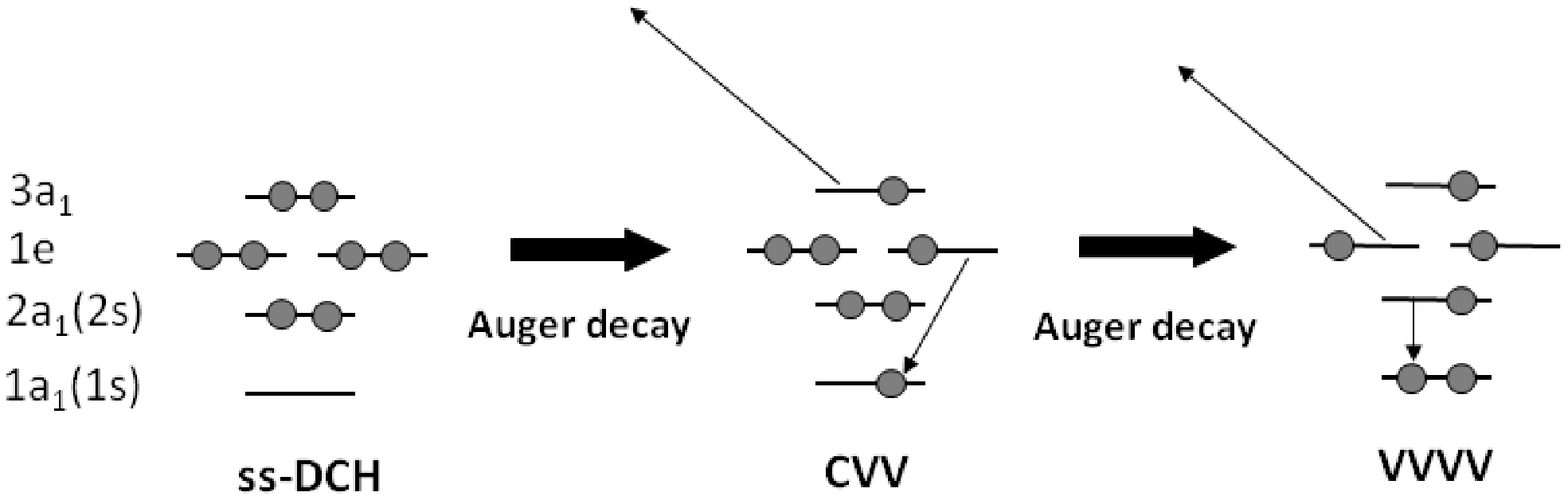}%
 \caption{\label{fig11}
 An example of two-step Auger decays originated from the ss-DCH state of NH$_3$. 
 Thin black arrows represent displacements of electrons during the Auger decays. 
 In the 1st Auger decay, the CVV state with one vacancy in the 3$a_1$ orbital 
 and another vacancy in the 1$e$ orbital is produced. Then in the 2nd Auger transition, 
 this CVV state decays to produce the VVVV state with additional vacancies in the 1$e$ 
 and 2$a_1$ orbitals.  
   }
\end{figure}

Auger decay of molecular DCH state mainly involves three different kinds of electronic 
states:\cite{ISI:000284407400011,ISI:000287357100010} DCH state, 
triply ionized states with one core hole and two valence holes (CVV states)
and quadruply valence ionized (VVVV) states. 
These electronic states and their relations are schematically shown in Fig. \ref{fig11}. 
In the first Auger transition, the DCH state decays to 
the CVV states, then these CVV states decay to the VVVV states in the second Auger transition.  
Thus, two Auger electrons are emitted in molecular DCH Auger decay, in contrast to one 
Auger electron emission in Auger decay of single core-hole state. 
Other processes may also be involved in molecular DCH decay, 
such as X-ray emission, 
double Auger process\cite{ISI:A19656228200002} where a valence electron fills the vacant core-hole 
with emission of two Auger electrons, 
or direct Auger decay from DCH state to VVVV state where two valence electrons 
simultaneously fill the vacant core-holes with an emission of an Auger electron. 
These processes are not treated in the present work, because 
the probability of X-ray emission is expected to be small for low-Z elements treated in the present 
work. Aso, description of the double Auger process and the direct Auger decay from DCH state to VVVV state 
requires the second order perturbation in terms of the Coulomb interaction,\cite{ISI:A1992HN47400050} 
and thus their transition probabilities are expected to be small compared to 
normal Auger process which can be described by the first order perturbation. 

In this work, Auger electron kinetic energy distributions are calculated and analyzed for 
the DCH states with K-shell vacancies of CH$_4$, NH$_3$ and H$_2$CO molecules. 
Auger decay in CH$_4$ and NH$_3$ involves only ss-DCH state.  
Auger decay in H$_2$CO, on the other hand, involves both ss- and ts-DCH states and therefore 
more complex Auger spectrum is expected. Although number of molecules is limited, 
our results may provide useful insights to discuss Auger decays of DCH states 
in other molecules. 

We employ the complete active space self-consistent field (CASSCF) and configuration 
interaction (CASCI) methods to evaluate energies of DCH, CVV and VVVV states.  
The experimental ionization energies of the ss-DCH states have been 
measured for several molecules\cite{ISI:000284407400011,ISI:000287357100010} 
and have been well reproduced by the CASSCF method.
\cite{ISI:000277756500014,ISI:000284407400011} 
Also, the CVV and VVVV energies of NH$_3$ obtained by the CASSCF and CASCI 
calculations roughly coincide with the peak positions in the experimental DCH Auger 
spectrum.\cite{ISI:000284407400011} 
%
In order to obtain theoretical Auger electron kinetic energy distribution, 
we have to calculate Auger intensiies other than the energies of the DCH, CVV and VVVV states. 
In the present work, we directly evaluate Auger amplitude in the Wentzel's formula, 
where the initial and the final states are represented by 
multi-configuratinal wave functions. 
As we will show in section II, the Auger amplitude is represented by a linear combination of 
two-electron integrals with core-hole, valence and continuum Auger electron orbitals. 
In the literature of molecular single core-hole (SCH) Auger decay, 
explicit evaluation of continuum Auger orbital has been limited to a few theoretical works, 
\cite{ISI:A1979GX19900030,ISI:A1982NY13400015,ISI:A1989AB70100016,ISI:A1992GY51800043}  
since treatment of such orbital in molecular system is rather difficult. 
Instead, several approximate methods have been introduced to avoid this difficulty. 
For instance, the molecular orbital of Auger electron is replaced by an continuum Auger 
orbital in atomic system.\cite{ISI:A1975AS23900009,ISI:A1995UC82800045} 
The Stieljes imaging method was also employed \cite{ISI:A1987F987200009} to represent 
continuum orbital in L$^2$ basis set.  
In other case, atomic population of valence orbitals is utilized to estimate 
absolute or relative Auger transition intensities,
\cite{ISI:A1979HA06900058,ISI:A1991ET47300059,ISI:A1994NV16900015,ISI:000181521100001} 
in which continuum orbital is not evaluated at all. 
In this work, the intensities of DCH $\to$ CVV and CVV $\to$ VVVV Auger decays are 
approximated by using atomic population of valence orbitals, 
which was originally introduced for molecular SCH Auger decay by 
Mitani et al.\cite{ISI:000181521100001} 
Although this approximation is not so accurate, it will be suitable to survey 
characteristic feature of Auger electron spectrum originated from molecular DCH Auger decay.

%
%





\section{Theoretical method}

\subsection{Auger Intensity}

Kinetic energies of Auger electrons can be evaluated using standard 
quantum chemistry method for bound electronic state. In the present work, 
we employ the CASSCF and CASCI methods to obtain Auger electron energies.
Estimation of Auger transition intensity, on the other hand, is not 
so straitforward as in evaluation of Auger electron kinetic energy. 
As in other theoretical works in molecular core-hole decay,
\cite{ISI:A1989AB70100016,ISI:A1992GY51800043,ISI:A1992MG04500001,ISI:A1994NV16900015,ISI:000181521100001} 
intensities of DCH $\to$ CVV Auger decay and subsequent CVV $\to$ VVVV Auger decay 
are evaluated by the Wentzel's formula,\cite{wentzel1927}
\begin{equation}
I_{fi} = 2\pi \left|\langle \Psi_{f} | \hat{H} - E | \Psi_{i} \rangle \right|^2 
\equiv 2\pi \left| t \right|^2,
\label{wentzel}
\end{equation}
where $\Psi_{i}$ and $\Psi_{f}$ are the wave functions for the initial and the final state, 
$\hat{H}$ is the Hamiltonian, $E$ is the energy of the initial state, and $t$ is 
the amplitude of the Auger transition. Atomic units are used in this expression. 
The wave function for the initial core-hole state, $\Psi_{i}$, is represented by a single 
or multi configurational $N$ electron function with bound molecular orbitals (MOs), 
where $N$ is the number of electrons in the initial state. In contrast, the wave function 
for the final state is in general represented as 
$\Psi_{f} = A \sum_a \Phi_{a} \phi_a$, where $\Phi_{a}$ is $N-1$ electron wave function for 
the final ion state $a$, $\phi_a$ represents continuum function for Auger electron in 
channel $a$, and $A$ is antisymmetlization operator. In this work, we employ single-channel 
expression of the final state $\Psi_{f}$ in which we include only one final ion state $\Phi_{a}$. 
When we calculate an intensity of Auger transition between the initial state $\Psi_{i}$ and 
the final ion state $\Phi_{a}$, the final state wave function is just represented as 
$\Psi_{f} = A \Phi_{a} \phi_a$ under the single-channel expression. 
Further discussion on this subject can be found in Ref. \onlinecite{ISI:A1992MG04500001} and references 
therein. 
In the present work, we used DCH and CVV state for  
$\Psi_{i}$ and $\Phi_{a}$, respectively, in calculation of 
DCH $\to$ CVV Auger intensity. 
Similarly, we used CVV and VVVV state for $\Psi_{i}$ and $\Phi_{a}$, respectively, 
in calculation of CVV $\to$ VVVV Auger intensity. 
%


Before going into detail of actual Auger intensity calculation, 
we show below that convenient short expressions are available for DCH $\to$ CVV Auger amplitude, 
when the initial DCH state and the final CVV ion state are represented by 
single configurational state functions (CSFs) and the frozen orbital approximation is adopted.  
Although these short expressions were not used to calculate Auger intensity in the present work, 
we used them to interpret and analyze our Auger spectra. 
In case of SCH Auger decay, it is well known that the Auger amplitude $t$ in Eq. (\ref{wentzel}) 
has convenient short expressions when the initial and the final ion states are represented 
by single 
CSFs.\cite{ISI:A1975AS23900009,ISI:A1990DC15100011,ISI:A1992MG04500001,ISI:000181521100001}
Assuming that the doublet SCH state is prepared by inner-shell ionization of 
closed-shell molecule, the Auger amplitude for the singlet final ion state with two vacancies 
in the $v$ and $w$ valence MOs is, 
\begin{eqnarray}
t &=& \sqrt{\frac{1}{2}} \left[ \left( kv|cw \right) + \left( kw|cv \right) \right]
  \hspace{10pt} \left( v \neq w \right),\\
  &=&  \left( kv|cv \right)  \hspace{10pt} \left( v = w \right).  
\label{sch1}
\end{eqnarray}
For triplet final ion state, the amplitude is 
\begin{equation}
t = \sqrt{\frac{3}{2}} \left[ \left( kv|cw \right) - \left( kw|cv \right) \right]. 
\label{sch3}
\end{equation}
Here $c$ and $k$ represent MOs for the core-hole vacancy and 
the continuum Auger electron, respectively. 
The expression $(ij|kl)$ represents two-electron integral, 
\begin{equation}
\left( ij|kl \right) = \int dr_{1} dr_{2} \phi_{i} \left( r_{1} \right)
\phi_{j} \left( r_{1} \right) \frac{1}{r_{12}} 
\phi_{k} \left( r_{2} \right)
\phi_{l} \left( r_{2} \right),
\label{def2eint}
\end{equation}
which involves MOs $\phi_{i}$, $\phi_{j}$, $\phi_{k}$ and $\phi_{l}$. 
In a similar manner, we can derive expression for amplitudes of Auger transition from DCH state 
to CVV state, assuming frozen orbital approximation as well as single CSF wave functions. 
The results are summarized in Table \ref{tab0}. 
We selected a specific spin-coupling scheme to represent doublet CVV states 
in Table \ref{tab0}, where intermediate spin state, singlet or triplet, was first formed in 
valence electrons, then it was coupled with doublet core electron to form total spin state with S=1/2. 
In other word, we represent two S=1/2 wave functions as 
$\Phi_{S} = 1/\sqrt{2} 
\left( \left|v \bar{w} c \right| - 
\left| \bar{v} w c \right| \right) $
and 
$\Phi_{T} = 1/\sqrt{6} 
\left( \left|v \bar{w} c \right| +  \left|\bar{v} w c \right| - 
2 \left| v w \bar{c} \right| \right) $.
Spin-coupling scheme is not unique when we construct total spin state from 
three S=1/2 particles.\cite{ISI:000181788600012} 
For example, the wave functions $\Phi_{S}$ and $\Phi_{T}$ shown above are constructed from 
spin coupling scheme $vw \cdot c$, which means that 
$v$ and $w$ orbitals form singlet or triplet configuration, then 
$c$ orbital couples this configuration. Similarly, other spin coupling 
scheme such as $v \cdot wc$ or $cv \cdot w$ is possible.
So, care must be taken when expressions in Table \ref{tab0} are compared with 
other calculation.  
The expressions in Table \ref{tab0} are very similar to those of the SCH Auger amplitudes, i.e., 
addition of two-electron integrals appears in the expression when valence electrons are 
coupled to form singlet, while subtraction of the integrals appears 
when valence electrons form triplet. 
From this observation, the Auger transition from DCH state to the CVV state with 
singlet intermediate valence spin state is expected to be more intense than  
the transition to the CVV state with triplet intermediate spin state, 
as in the case of SCH Auger decay. 
Although it may be possible to derive similar expressions for CVV $\to$ VVVV Auger decay, 
we did not attempt it because there are too many combinations in orbitals. 
In the present work, we directly evaluated Eq. (\ref{wentzel}) using CI wave functions 
of CVV and VVVV states, as will be described below. 
Although the equations in Table \ref{tab0} were not used in our actual calculation, 
they may be usefull in other theoretical or experimental work when quick evaluation of 
DCH $\to$ CVV Auger intensity is required. 

When the initial and the final ion states are represented by more general 
multi configurational functions, the Auger amplitude is represented as
\begin{equation}
t = \sum_{v,w} C_{v,w} \left( kv|cw \right),
\label{ciamp}
\end{equation} 
where the summation of MOs $v$ and $w$ is taken over all active valence orbitals.  
The coefficients $C_{v,w}$ depend on CI coefficients of the wave functions, 
determinant-CSF conversion coefficients, and relative ordering of MOs in 
determinants of the initial and final states. As in the case of single CSF wave functions, 
frozen orbital approximation was assumed in Eq. (\ref{ciamp}). 
We employ this frozen orbital approximation because of its computational 
efficiency. If we used non-orthogonal orbital sets, considering relaxation effect, 
then evaluation of Auger amplitude would be much difficult because of long calculation time.
Discussion on validity of the frozen orbital approximation for single core-hole Auger decay  
can be found in Ref. \onlinecite{ISI:A1992MG04500001} and references therein.
In order to obtain Auger spectrum, we directly evaluated the Auger amplitudes $t$ in 
Eqs. (\ref{wentzel}) and (\ref{ciamp}) in combination with multi-configurational expressions 
(CI wave functions) of the initial and the final states. 
Eqs.(\ref{wentzel}) and (\ref{ciamp}) can be used for any type of Auger transition involving  
one core-hole decay, two-valence hole creation and an emission of one Auger electron, 
which includes DCH$\to$CVV and CVV$\to$VVVV Auger decays as well as SCH$\to$VV Auger decay.
It is important to recognize that $C_{v,w}$ in Eq. (\ref{ciamp}) contains the product of the CI 
coefficients of the initial and final ion state wave functions, thus 
only the CSFs with large CI coefficients contribute to the Auger amplitude $t$. 

The two-electron integral $(kv|cw)$ in Eqs. (\ref{sch1}),(\ref{sch3}) and (\ref{ciamp}) 
contains MO $k$ for the continuum Auger electron. 
We used the method described in Mitani et al.\cite{ISI:000181521100001} which 
approximates this two-electron integral $\left( kv|cw \right)$ by the population of 
MO $w$ on the atom A where the core-hole orbital $c$ is localized. 
Although this approximation is not so accurate, it can reproduce qualitative 
feature of the experimental SCH Auger spectra of H$_2$O and NH$_3$,\cite{ISI:000181521100001} 
and will be appropriate to understand qualitative nature of DCH Auger spectrum as well.

\subsection{Detail of the calculation}

The energies of DCH, CVV and VVVV states as well as Auger intensities 
were evaluated by fixed bond length calculations.  
Since the lifetime of DCH and CVV states are expected to be similar or 
shorter than the lifetime of SCH state, e.g., less than 10 fs for C, N or O elements,\cite{ISI:A1991FV18400032} 
the effect of the vibrational motion may be small. 
Experimental geometries of CH$_4$, NH$_3$ and H$_2$CO molecules\cite{ChemElem2nd,GR79,OK60}  
were used in this work. 
Calculation for H$_2$CO molecule was performed with $C_{2v}$ point group symmetry. 
Although CH$_4$ and NH$_3$ molecules belong to $T_d$ and $C_{3v}$ point group 
symmetries, respectively, we used $D_2$ and $C_s$ symmetries in the calculations. 
The energies and CI coefficients of the CVV and VVVV states were obtained by 
the CASCI calculations using the MOs prepared  
by the state-averaged (SA) CASSCF \cite{ISI:A1985AGJ3300004,ISI:A1985AJG5000041} calculations  
for the low-lying CVV and VVVV states. 
For CH$_4$, 60 CVV and 300 VVVV states were obtained by the CASCI calculations 
based on the CASSCF calculations for the low-lying 12 CVV and 10 VVVV states, respectively.  
For NH$_3$, 100 CVV states were calculated by the CASCI method 
based on the CASSCF calculation for the low-lying 15 CVV states, 
and 300 VVVV states were obtained by the CASCI method using 
the MOs taken from the CASSCF calculation for the low-lying 11 VVVV states. 
For H$_2$CO, 1500 C1s$^{-1}$ CVV, 1500 O1s$^{-1}$ CVV and 3000 VVVV states 
were obtained by the CASCI calculations based on the CASSCF calculations 
for the low-lying 20 C1s$^{-1}$ CVV, 24 O1s$^{-1}$ CVV and 30 VVVV states, respectively. 
Note that these numbers refer to total numbers of calculated CVV or VVVV states 
in the molecules, i.e., summation of calculated states over all irreducible representations. 
These electronic states obtained by the CASCI calculations were sufficient to 
cover whole energy range of the DCH $\to$ CVV as well as the CVV $\to$ VVVV Auger spectra of 
CH$_4$, NH$_3$ and H$_2$CO.  
The cc-pVTZ basis set \cite{1989JChPh..90.1007D} was employed for all these calculations.  
We used frozen Hartree-Fock orbitals to represent 1s core orbitals, while the other orbitals 
were fully relaxed and optimized in the CASSCF calculations. 
As in our previous works on molecular DCH states,\cite{ISI:000277756500014,ISI:000281296900043} 
all valence electrons were distributed in the active orbital space composed of all 
available valence MOs, while occupation of core-hole orbital was 
explicitly restricted. 
The same set of active orbitals was used in both the CASSCF and CASCI calculations. 
The total number of active orbitals is 8, 7 and 10 for CH$_4$, NH$_3$ and H$_2$CO, 
respectively. 
With these active orbitals, typical number of configurations 
in the CVV states is 670, 550 and 12500 for CH$_4$, NH$_3$ and H$_2$CO, respectively. 
For the VVVV states, the number of CSFs is 90, 100 and 3500 for CH$_4$, NH$_3$ and H$_2$CO, 
respectively.
The energies of the DCH states for CH$_4$, NH$_3$ and H$_2$CO 
were also obtained by the CASSCF calculations, using the same basis set and 
active orbitals as in the calculations for the CVV and VVVV states. 
Our CASSCF double ionization energies (DIEs) are 650.2, 891.0, 658.0 and 1168.5 eV for 
the CH$_4$ C1s$^{-2}$, NH$_3$ N1s$^{-2}$, 
H$_2$CO C1s$^{-2}$, and H$_2$CO O1s$^{-2}$ ss-DCH states, respectively. 
The DIE of the H$_2$CO C1s$^{-1}$O1s$^{-1}$ ts-DCH state is 846.6 eV for the singlet, 
and 847.0 eV for the triplet state.  
These DIEs were used to obtain the Auger kinetic energies of the DCH $\to$ CVV Auger decays. 
For these CASSCF and CASCI calculations, molpro program package \cite{MOLPRO} was mostly used. 
In addition, congen and scatci modules in the UK R-matrix codes \cite{Mo98} were partly used to 
analyze the electronic states. 
%


%
In evaluation of the Auger intensities, we used the L\"owdin atomic 
population of the MOs which were taken from the CASSCF calculation 
for the CVV states. The population obtained by the MOs for the VVVV states gave similar Auger 
intensities. Although choice of population, L\"owdin or Mulliken, affects Auger intensities  
as demonstrated by Mitani et al.,\cite{ISI:000181521100001} 
qualitative feature of the result does not change much. 
In Auger intensity evaluation using Eqs. (\ref{wentzel}) and (\ref{ciamp}), 
wave function of DCH state was approximated by a single CSF, since 
weights of the other configurations were small. 
For CVV and VVVV states, five configurations with the largest CI coefficients 
were taken into account in the Auger intensity evaluation.
We performed a test calculation on CH$_4$ DCH Auger spectra, taking into account 
30 configurations with the largest CI coefficients. The results obtained with 30 and 5 
configurations are almost the same, as shown in Fig. S1.  
This indicates that inclusion of five configurations is enough in Auger intensity evaluation. 
The intensities of the 1st Auger transitions were normalized as 
$\sum_{i} I \left( {\rm DCH}, {\rm CVV}_{i} \right)$ = 1 or 3, where the summation is 1 for 
the singlet ss- and ts-DCH states and 3 for the triplet ts-DCH state, 
and $I \left( {\rm DCH}, {\rm CVV}_{i} \right)$ is the intensity for transition from the DCH state 
to the final state with the $i$th CVV ion state. 
The normalization for the 2nd Auger intensities is 
$\sum_{j} I \left( {\rm CVV}_{i}, {\rm VVVV}_{j} \right)$ = $I \left( {\rm DCH}, {\rm CVV}_{i} \right)$, 
where $I \left( {\rm CVV}_{i}, {\rm VVVV}_{j} \right)$ is the intensity for the transition from 
the $i$th CVV state to the final state with the $j$th VVVV ion state. 
This expression for normalization of the 2nd Auger intensity indicates that two conditions 
should be satisfied at once in order to have large intensity in the 2nd Auger decay; 
(1) The $i$th CVV state is populated enough in the 1st Auger decay, or, 
$I \left( {\rm DCH}, {\rm CVV}_{i} \right)$ is large enough, and  
(2) The $i$th CVV state and the $j$th VVVV ion state have enough transition moment 
in Eq. (\ref{wentzel}), or, $I \left( {\rm CVV}_{i}, {\rm VVVV}_{j} \right)$ is 
large enough. These two conditions severely restrict possible pair of CVV and VVVV 
states which has noticeably large intensity in the 2nd Auger decay. 

\section{Results and Discussion}

\subsection{CH$_{4}$}

\begin{figure}
 \includegraphics[scale=2.0]{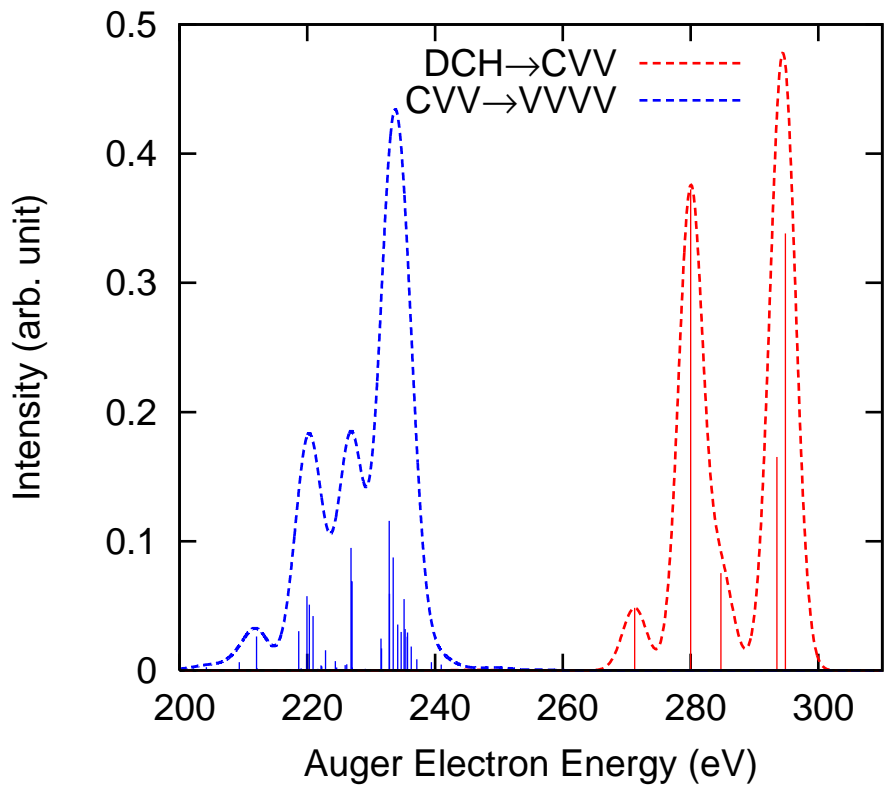}%
 \caption{\label{fig1}
 Calculated intensities of the Auger decays of the CH$_4$ C1s$^{-2}$ 
 DCH state (DCH$\to$CVV) and the C1s$^{-1}$ CVV states (CVV$\to$VVVV), 
 as a function of Auger electron kinetic energy. 
 The vertical full lines represent the discrete Auger spectrum obtained 
 by the CASCI wave functions with the frozen orbital approximation. 
 The dashed lines are obtained by convoluting the discrete Auger intensities 
 with Gaussian function having 4.5 eV width. 
   }
\end{figure}

In Fig. \ref{fig1}, calculated CH$_4$ Auger intensities 
are shown as a function of Auger electron kinetic energy,  
where the 1st Auger transitions from the C1s$^{-2}$ DCH state to the 
C1s$^{-1}$ CVV states and the 2nd Auger transitions from the C1s$^{-1}$ CVV states 
to the VVVV states contribute the spectrum. 
Since the main configuration of the CH$_4$ DCH state is $(1s)^0(2a_1)^2(1t_2)^6$, 
the $2a_1$ and $1t_2$ valence orbitals mainly participate to the Auger transitions.  
Main part of the DCH $\to$ CVV Auger spectrum extends from 270 to 290 eV, and 
the CVV $\to$ VVVV Auger spectrum extends from 210 to 240 eV. 
These two different Auger spectra are well separated in energy each other. 
Number of discrete Auger transitions, indicated as vertical bars in Fig. \ref{fig1}, 
is 6 for the DCH $\to$ CVV Auger decay and 300 for the CVV $\to$ VVVV Auger decay. 
The convoluted DCH $\to$ CVV Auger spectrum in Fig. \ref{fig1}, 
obtained by convolution of the discrete Auger spectrum with Gaussian having 4.5 eV width, 
has 3 large distinct peaks, whereas the convoluted CVV $\to$ VVVV Auger spectrum has 4 peaks. 
The highest energy peak ($\sim$ 295 eV) in the convoluted DCH $\to$ CVV Auger spectrum 
corresponds to the $(t_2)^{-2}$ valence vacancy creation. 
The second highest ($\sim$ 280 eV) and the 3rd highest ($\sim$ 270 eV) energy peaks 
in the DCH $\to$ CVV spectrum are formed by the $(2a_1)^{-1}(t_2)^{-1}$ and $(2a_1)^{-2}$ 
valence hole creations, respectively. 
The highest ($\sim$ 235 eV) and the lowest ($\sim$ 210 eV) energy peaks in the CVV $\to$ 
VVVV Auger spectrum originate from the $(t_2)^{-2}$ and $(2a_1)^{-2}$ vacancy creations, 
as in the DCH $\to$ CVV Auger case. 
Two peaks in the middle of the CVV $\to$ VVVV Auger spectrum are formed by 
the $(2a_1)^{-1}(t_2)^{-1}$ vacancy creation, where  
the initial states of the higher energy peak ($\sim$ 225 eV) have the $(2a_1)^{1}$ 
configurations and those of the lower energy peak ($\sim$ 220 eV) have the $(2a_1)^{2}$ 
configurations. 

\begin{figure}
 \includegraphics[scale=1.5]{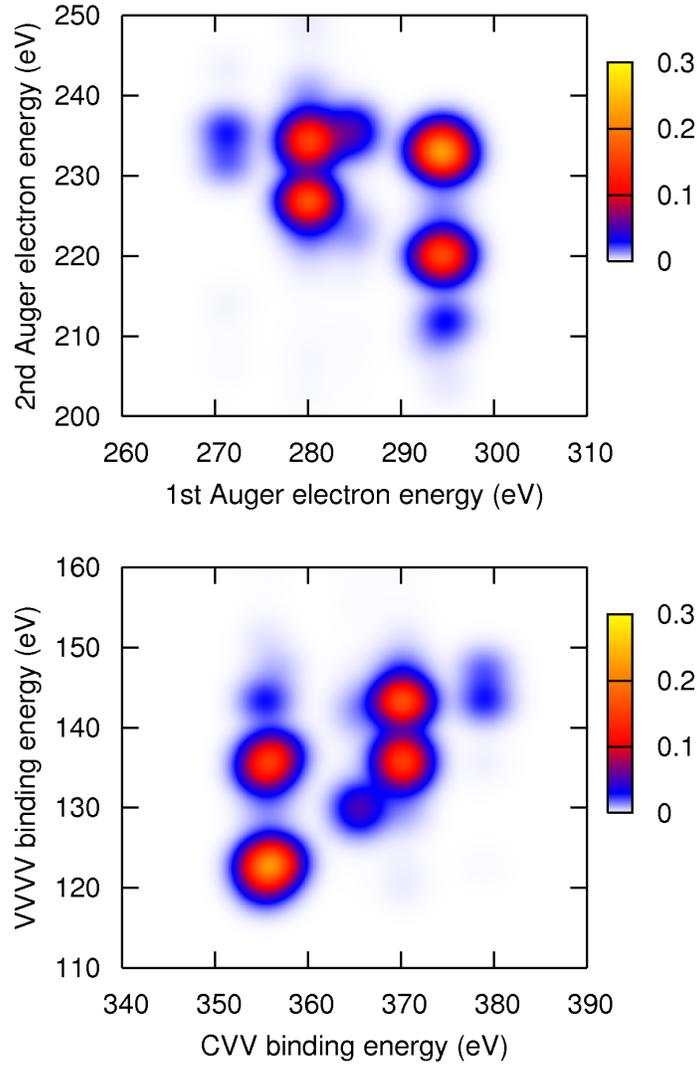}%
 \caption{\label{fig2} 
 (Upper panel) Two-dimensional (2D) intensity of the two-step Auger decays of the 
 CH$_4$ C1s$^{-2}$ ss-DCH state, as functions of kinetic energies of the 1st and the 2nd Auger electrons. 
 The 1st electrons are emitted by the Auger transition from the C1s$^{-2}$ DCH state to 
 the C1s$^{-1}$ CVV states, and the 2nd electrons are emitted by the transition 
 from the C1s$^{-1}$ CVV states to the VVVV states. 
 This 2D spectrum represents a probability of finding two Auger electrons 
 at specific pair of energies. 
 (Lower panel) 2D Auger intensity as functions of CVV and VVVV binding energies, 
 converted from the 2D spectrum in the upper panel.  
 These intensities were obtained by smoothing calculated discrete Auger intensities, 
 as in the convoluted intensities in Fig. \ref{fig1}. 
 Unit of the intensity is arbitrary.  
   }
\end{figure}

The kinetic energies of the 1st and 2nd Auger electrons have correlation 
because these two Auger decays are not independent as shown schematically in Fig. \ref{fig11}. 
The upper panel of Fig. \ref{fig2} shows 2D Auger intensity 
distribution as functions of the 1st and 2nd Auger electron kinetic energies, 
obtained by smoothing theoretical discrete Auger spectrum.  
Integration of this 2D Auger spectrum along the vertical axis (the 2nd Auger electron energy) 
gives the convoluted DCH $\to$ CVV Auger spectrum in Fig. \ref{fig1}, and integration along 
the horizontal direction (the 1st Auger electron energy) gives the CVV $\to$ VVVV Auger 
spectrum. 
In the upper panel of Fig. \ref{fig2}, several distinct and round-shaped high-intensity regions 
are recognized. 
The highest intensity region around the 1st Auger energy of 295 eV and the 2nd 
Auger energy of 235 eV corresponds to the $(1t_2)^{-2}$ valence hole creation followed 
by another $(1t_2)^{-2}$ valence hole creation
in the successive two Auger transitions. Next to this highest intensity peak, 
there are three peaks with the 2nd highest Auger intensities. 
The peak around the 1st Auger energy of 295 eV and the 2nd Auger energy of 220 eV 
corresponds to the $(1t_2)^{-2}$ vacancy creation in the 1st Auger decay, followed 
by the $(2a_1)^{-1}(1t_2)^{-1}$ vacancy creation in the 2nd Auger decay. 
The peak at the 1st Auger energy of 280 eV and the 2nd Auger energy of 235 eV 
corresponds to successive formation of the $(2a_1)^{-1}(1t_2)^{-1}$ and the 
$(1t_2)^{-2}$ valence vacancies. 
The peak around the 1st Auger energy of 280 eV with the 2nd Auger energy of 225 eV 
is formed by the $(2a_1)^{-1}(1t_2)^{-1}$ vacancy creation followed by another 
$(2a_1)^{-1}(1t_2)^{-1}$ vacancy creation. 
Note that Lablanquie et al. reported the experimental 2D Auger spectrum for 
the ss-DCH Auger decay of N$_2$ molecule,\cite{ISI:000287357100010} 
by measuring kinetic energies of the two Auger electrons.  
So far, such experimental 2D Auger spectrum is not available for the other molecules. 

\begin{figure}
 \includegraphics[scale=1.5]{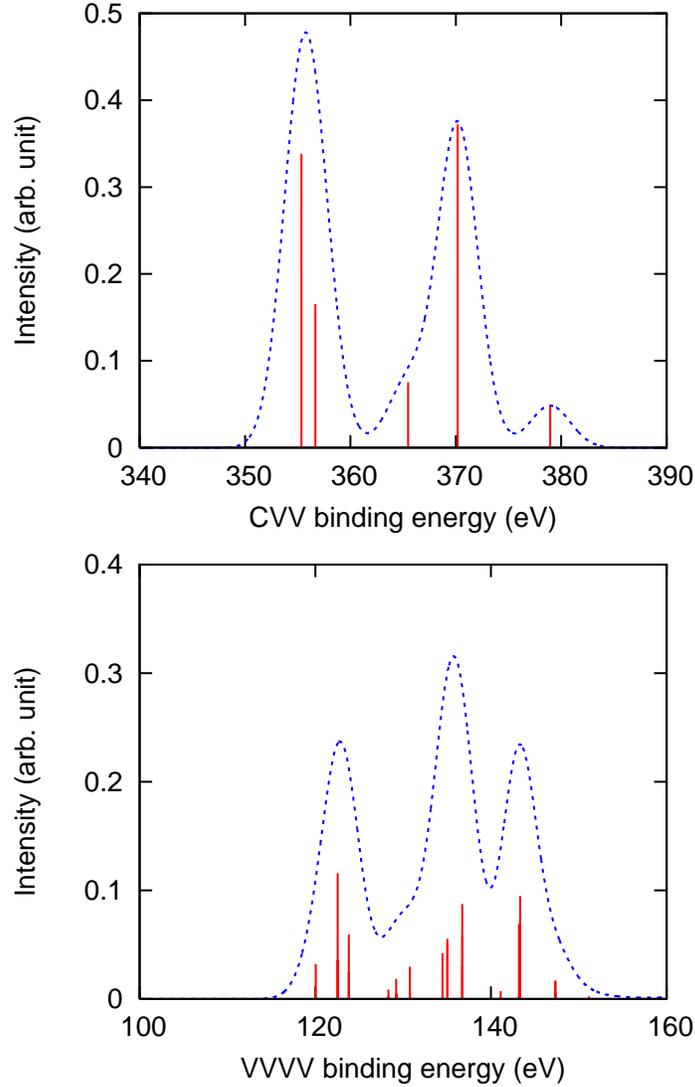}%
 \caption{\label{fig3} 
 Integrated 1D Auger intensities as a function of CVV binding energy (upper panel) 
 and VVVV binding energy (lower panel). 
 These spectra can be obtained by integrating the 2D Auger spectrum in the lower panel 
 of Fig. \ref{fig2} by VVVV energy (upper panel) or CVV energy (lower panel). 
 The vertical bars represent original theoretical intensities which can be directly 
 obtained from our calculation. 
 The other details are the same as in Fig. \ref{fig1}.  
   }
\end{figure}

Binding energies of DCH state ($E_{\rm DCH}$), CVV state ($E_{\rm CVV}$) and 
VVVV state ($E_{\rm VVVV}$) are related to the 1st and 2nd Auger electron kinetic 
energies ($E_{\rm Auger1}$ and $E_{\rm Auger2}$) as, 
$E_{\rm DCH} = E_{\rm CVV} + E_{\rm Auger1}$ 
and $E_{\rm CVV} = E_{\rm VVVV} + E_{\rm Auger2}$. 
Using these relations as well as experimental or 
theoretical value of $E_{\rm DCH}$, the 2D Auger spectrum in the upper panel of 
Fig. \ref{fig2} can be converted to the 2D Auger spectrum as functions of 
CVV and VVVV binding energies, as shown in the lower panel of Fig. \ref{fig2}. 
In this plot, contributions of CVV and VVVV states to the Auger intensity 
can be easily recognized: 
the highest intensity peak at bottom left corner of the plot is 
related to the lowest energy CVV and VVVV states. for example. 
By integrating this 2D Auger spectrum along the horizontal axis (CVV energy), 
integrated one-dimensional (1D) Auger intensity as a function of VVVV binding energy 
can be obtained as shown in the lower panel of Fig. \ref{fig3}.  
Similarly, integration along the vertical axis (VVVV energy) gives integrated 1D 
Auger intensity as a function of CVV binding energy as shown in the upper panel of 
Fig. \ref{fig3}. 
Although we need integration to extract these 1D spectra from 
experimental data, theoretical 1D spectra can be obtained directly from our calculation, 
without explicit integration. The original discrete spectra are shown as vertical bars in Fig. \ref{fig3}. 
Three distinct peaks are seen in both panels, with the highest energy peak 
corresponds to the $(2a_1)^0(1t_2)^{n}$ configuration, the second highest peak 
corresponds to the $(2a_1)^1(1t_2)^{n-1}$ configuration
and the lowest energy peak corresponds to the $(2a_1)^2(1t_2)^{n-2}$ configuration, 
where $n$ is 6 for the CVV states in the upper panel and 4 for the VVVV states in 
the lower panel.

For reference, ionization energies of the calculated CVV states are summarized  
in Table \ref{tab1} with their main configurations and intensities for 
the 1st Auger decay. 
The VVVV states are not provided because there are too many states. 
Because of spin conservation in Eq. (\ref{wentzel}), the singlet ss-DCH state of 
CH$_4$ decays only to the doublet CVV states. 
When the wave functions of the DCH and CVV states are approximated by a single CSF, 
expressions of the ss-DCH $\to$ CVV Auger intensities can be classified into three different 
types as described in Sec. II.A and Table \ref{tab0}. 
According to this classification, Auger intensity tends to be larger for the CVV final ion 
state with singlet intermediate spin state in valence electrons than that with 
triplet intermediate spin state in valence electrons. 
The calculated Auger intensities in Table \ref{tab1} roughly obey this trend. 

\subsection{NH$_{3}$}

\begin{figure}
 \includegraphics[scale=2.0]{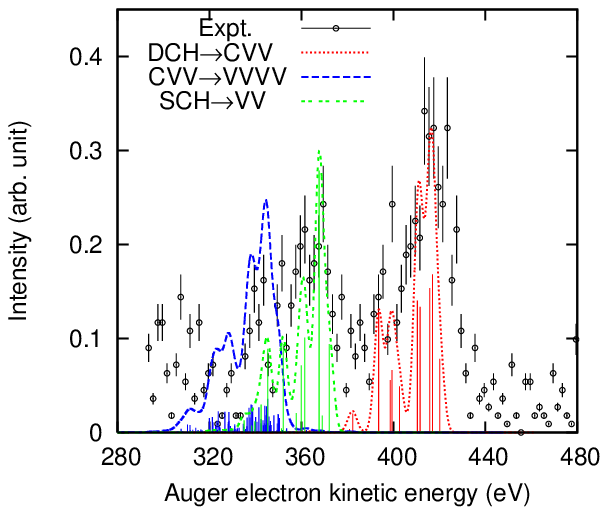}%
 \caption{\label{fig4} 
 Calculated and experimental Auger intensities of the NH$_3$ core-hole decays as a function of 
 Auger electron kinetic energy.  
 DCH$\to$CVV: the transition from the N1s$^{-2}$ DCH state to the N1s$^{-1}$ CVV states, 
 CVV$\to$VVVV: the transition from the N1s$^{-1}$ CVV states to the VVVV states, 
 SCH$\to$VV: the Auger decay of the N1s$^{-1}$ SCH state.
 Expt.:experimental Auger spectrum of Eland et al.\cite{ISI:000284407400011}. 
 The other details are the same as in Fig. \ref{fig1}. 
   }
\end{figure}

In Fig. \ref{fig4}, calculated Auger spectrum 
of NH$_3$ DCH decay is shown as a function of Auger electron kinetic energy, 
which include contributions from the 1st Auger transitions from the N1s$^{-2}$ DCH state 
to the N1s$^{-1}$ CVV states and the 2nd Auger transitions from the N1s$^{-1}$ CVV states 
to the VVVV states. 
For comparison, experimental NH$_3$ DCH Auger spectrum of Eland et al.
\cite{ISI:000284407400011} 
and theoretical Auger spectrum of NH$_3$ SCH $\to$ VV (doubly valence ionized states) 
Auger decay are also shown in the figure. 
The SCH Auger spectrum was calculated by the same procedure as we used for the DCH $\to$ CVV 
and CVV $\to$ VVVV Auger decays. 

Main part of the theoretical DCH $\to$ CVV Auger spectrum extends from 380 to 420 eV, 
whereas the CVV $\to$ VVVV and SCH $\to$ VV Auger spectra extend from 300 to 350 eV 
and 335 to 375 eV, respectively. 
Although the CVV $\to$ VVVV and SCH $\to$ VV Auger spectra overlap partly, the main part of the 
DCH $\to$ CVV Auger spectrum is well separated from the CVV $\to$ VVVV and SCH $\to$ VV 
spectra. 
Number of discrete Auger transitions, indicated as vertical bars in Fig. \ref{fig4}, 
is 14 for the DCH $\to$ CVV Auger decay, 23 for the SCH $\to$ VV Auger decay, and 995 
for the CVV $\to$ VVVV Auger decay. 
Because of difference in number of transitions, discrete spectra of the DCH $\to$ CVV and 
CVV $\to$ VVVV Auger decays look rather different in Fig. \ref{fig4}. 
However, when these spectra are convoluted by Gaussian function, close similarity is observed 
between these two Auger spectra. Also, overall shape of the convoluted SCH $\to$ VV Auger 
spectrum resembles those of the convoluted DCH $\to$ CVV and CVV $\to$ VVVV Auger spectra. 
Compared to the position of the convoluted SCH $\to$ VV Auger spectrum, 
the DCH $\to$ CVV Auger spectrum is located about 50 eV higher in energy, 
and the CVV $\to$ VVVV spectrum is located about 20 eV lower in energy. 
Origin of the energy shift of the DCH $\to$ CVV Auger spectrum relative to 
the SCH $\to$ VV spectrum is mainly attributed to the Coulomb repulsion of the core holes, 
which increases the energy of the DCH state.\cite{ISI:A1986F086500030,ISI:000277756500014} 
In addition, the CVV states are stabilized by existence of two valence holes 
which weaken the Coulomb screening of the nuclear charges. 
This stabilization contributes to the shift of the DCH $\to$ CVV Auger spectrum 
into the higher energy. The shift of the CVV $\to$ VVVV Auger spectrum 
to the lower energy can also be explained by this stabilization 
of the CVV states. 
Each convoluted Auger spectrum has 5 distinct peaks, which can be 
assigned in terms of the occupied valence MOs in the main configurations 
of the DCH and SCH states: the $2a_1$, $1e$ and $3a_1$ orbitals.   
The highest energy peak has the largest 
intensity, accompanying a faint shoulder structure at higher energy side. 
This shoulder structure corresponds to the Auger transition with $(3a_1)^{-2}$ valence 
vacancy creation, whereas the main peak corresponds to the $(1e)^{-1}(3a_1)^{-1}$ vacancy 
creation. The second highest energy peak, with the second largest intensity, originates  
from the $(1e)^{-2}$ vacancy creation. The 3rd, 4th and 5th peaks are formed 
by the $(2a_1)^{-1}(3a_1)^{-1}$, $(2a_1)^{-1}(1e)^{-1}$ and $(2a_1)^{-2}$ valence hole 
creations, respectively. The intensity of the 5th peak is much smaller than the other peaks. 

The experimental spectrum of Eland et al.,\cite{ISI:000284407400011} obtained by 
the triple electron coincidence method with synchrotron radiation, is compared with 
our results in Fig. \ref{fig4}.
Although the experimental spectrum was plotted as a function of CVV binding energy 
in Ref. \onlinecite{ISI:000284407400011}, we converted the spectrum as a function of 
Auger electron kinetic energy using the experimental NH$_3$ DCH binding energy of 892 eV. 
The large peak around 425 eV in the experimental spectrum coincides well 
with the $(3a_1)^{-2}$, $(1e)^{-1}(3a_1)^{-1}$ and $(1e)^{-2}$ peaks in the  
calculated DCH $\to$ CVV Auger spectrum. The rise of experimental intensities 
around 400 eV may have relation with the $(2a_1)^{-1}(3a_1)^{-1}$ and 
$(2a_1)^{-1}(1e)^{-1}$ peaks in the calculated spectrum. 
The experimental intensity appears to increase at around 380 eV, where  
the calculated $(2a_1)^{-2}$ peak exists. However, association of this 380 eV 
peak with the calculated peak is not clear because of difference in intensities as well as 
experimental uncertainty. 
The experimental peaks around 360 - 370 eV coincide well with the calculated SCH $\to$ VV peaks 
with the $(3a_1)^{-2}$, $(1e)^{-1}(3a_1)^{-1}$ and $(1e)^{-2}$ vacancy creations. 
As noted by Eland et al.,\cite{ISI:000284407400011} the origin of these peaks around 
360 - 370 eV is secondary electrons created by the SCH $\to$ VV Auger electrons or 
photoelectrons of the SCH formation. Thus, the height of these peaks do not 
directly reflect the SCH formation cross section, which is much larger than the DCH formation 
cross section.\cite{ISI:000287357100010} Also, the shape of the SCH related peaks may be distorted from the original 
SCH Auger spectrum, because of the process involved in the secondary electron creation. 
We just adjusted the height of the calculated SCH Auger spectrum to the experimental peak 
around 370 eV. 
There is a peak around 340 eV in the experimental spectrum, whose 
location is very close to the calculated CVV $\to$ VVVV Auger peaks with the 
$(3a_1)^{-2}$, $(1e)^{-1}(3a_1)^{-1}$ and $(1e)^{-2}$ vacancy creations. 
Although intensities are smaller, the $(2a_1)^{-1}(3a_1)^{-1}$ and $(2a_1)^{-1}(1e)^{-1}$ 
peaks in the SCH $\to$ VV Auger spectrum may also have contribution to the experimental 
peak at 340 eV. 
Below 330 eV, association of the calculated CVV $\to$ VVVV Auger spectrum and the 
experimental spectrum is not clear. 
Although agreement with the experimental spectrum and our result 
looks modestly good on the whole, some degree of discrepancy exists, especially 
around the 2nd Auger component. Origin of the difference may be related to 
the approximations employed in our calculation, e.g., the frozen orbital 
approximation or neglect of the shake-up satellite DCH states. 
By considering orbital relaxation or the satellite states, 
the shape of the theoretical Auger spectrum may be changed. 
In addition, experimental noise from SCH signal can be another source of 
the discrepancy, which may be reduced by performing four-electron 
coincidence experiment. 

\begin{figure}
 \includegraphics[scale=1.5]{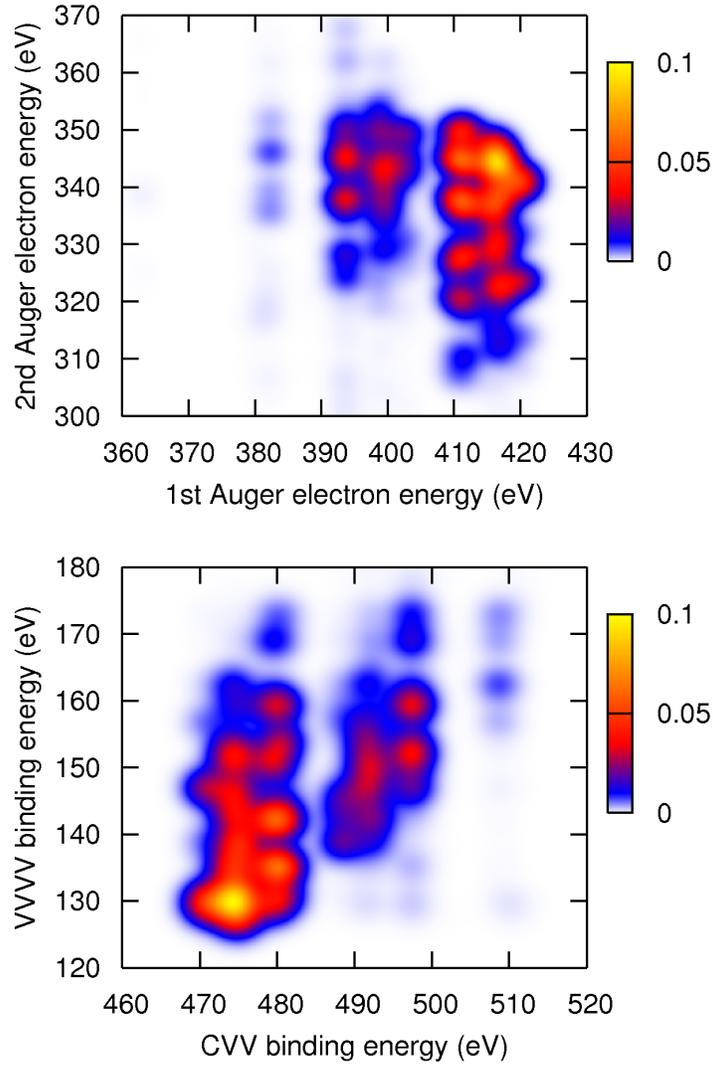}%
 \caption{\label{fig5} 
 (Upper panel) 2D intensity of the two-step Auger decays of the NH$_3$ N1s$^{-2}$ DCH state, 
 as functions of kinetic energies of the 1st and the 2nd Auger electrons. 
 (Lower panel) 2D Auger intensity as functions of CVV and 
 VVVV binding energies, converted from the 2D spectrum in the upper panel. 
 The other details are the same as in Fig. \ref{fig2}. 
   }
\end{figure}

In the upper panel of Fig. \ref{fig5}, 2D Auger spectrum 
is shown as functions of the 1st and the 2nd Auger kinetic energies.  
Compared to the 2D Auger intensity distribution for the CH$_4$ DCH decay in Fig. \ref{fig2}, 
high-intensity regions look less distinct in the NH$_3$ Auger spectrum. 
For both the 1st and 2nd Auger electrons, the higher kinetic energy corresponds to 
vacancy creation in the $3a_1$ and $1e$ orbitals, and 
the lower kinetic energy corresponds to vacancy creation in the $2a_1$ orbital. 
Thus, the highest-intensity region surrounded by the 1st Auger energies of 410 - 420 eV 
and the 2nd Auger energies of 340 - 350 eV mainly corresponds to 
the $(1e3a_1)^{-2}$ valence hole creation followed by another 
$(1e 3a_1)^{-2}$ valence hole creation in the sequential two-step Auger decays. 
The second highest intensity regions are located 
next to the highest intensity region, one surrounded by the 1st Auger energies of 410 - 420 eV 
and the 2nd Auger energies of 320 - 335 eV, and the other surrounded by the 1st Auger 
energies of 390 - 405 eV and the 2nd Auger energies of 340 - 350 eV. 
The former corresponds to the $(1e 3a_1)^{-2}$ valence hole creation followed 
by the $(2a_1)^{-1}(1e 3a_1)^{-1}$ valence hole creation, and  
the later corresponds to the $(2a_1)^{-1}(1e 3a_1)^{-1}$ valence hole creation followed 
by the $(1e 3a_1)^{-2}$ valence hole creation. 

\begin{figure}
 \includegraphics[scale=1.5]{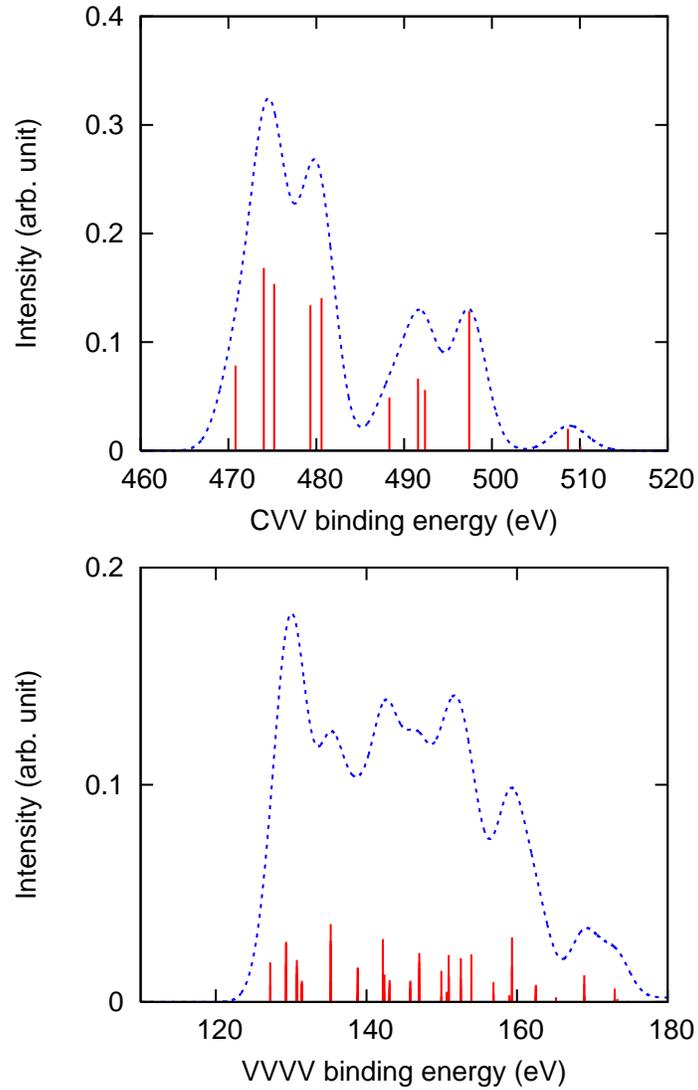}%
 \caption{\label{fig6} 
 Integrated 1D Auger intensities as a function of CVV binding energy (upper panel) and 
 VVVV binding energy (lower panel), obtained by integrating the 2D Auger spectrum 
 in the lower panel of Fig. \ref{fig5}. The other details are the same as in Fig. \ref{fig3}. 
   }
\end{figure}

As in the CH$_4$ case, the 2D Auger spectrum in the upper panel of Fig. \ref{fig5} 
can be converted to the 2D intensity as functions of CVV and VVVV binding energies 
as shown in the lower panel of Fig. \ref{fig5}. This plot shows clearly that the highest 
intensity region corresponds to the low-lying CVV and VVVV states. 
Integration of the 2D Auger spectrum in the lower panel of Fig. \ref{fig5} along 
the vertical or the horizontal axis gives the integrated 1D Auger spectra 
as shown in Fig. \ref{fig6}. The integrated Auger intensity as a function of CVV binding 
energy in the upper panel of Fig. \ref{fig6} has 3 large peaks which correspond to the CVV 
states with $(2a_1)^2$, $(2a_1)^1$ and $(2a_1)^0$ configurations. 
The integrated Auger intensity as a function of VVVV binding 
energy in the lower panel has more complicated structure compared to the intensity in 
the upper panel. We can still make assignment of these peaks in terms of electronic 
configuration, e.g., the lowest energy peak corresponds to the configuration 
$(2a_1)^{2}(1e)^{2}(3a_1)^{0}$, the second lowest peak to the configuration 
$(2a_1)^{2}(1e)^{1}(3a_1)^{1}$, 
while the peaks in the middle of the spectrum are related to the $(2a_1)^{1}$ type configuration.
Compared to the integrated 1D spectra of CH$_4$ VVVV states, the integrated 1D spectrum of NH$_3$ VVVV 
states is distributed broader and looks less structured. 
This difference is attributed to the different number of occupied valence orbitals 
in CH$_4$ (two orbitals) and NH$_3$ (three orbitals). 
This suggests that it may be hard for large molecule to recognize 
peak structure in integrated 1D spectra.

For reference, ionization energies of the calculated CVV states are listed 
in Table \ref{tab2} with their main configurations and the intensities for 
the DCH $\to$ CVV Auger decay.  
As in the CH$_4$ case, the DCH $\to$ CVV Auger intensity tends to be larger 
for the CVV final ion states with singlet intermediate valence spin than the ion states 
with triplet intermediate valence spin. 

\subsection{H$_{2}$CO}

\begin{figure}
 \includegraphics[scale=1.5]{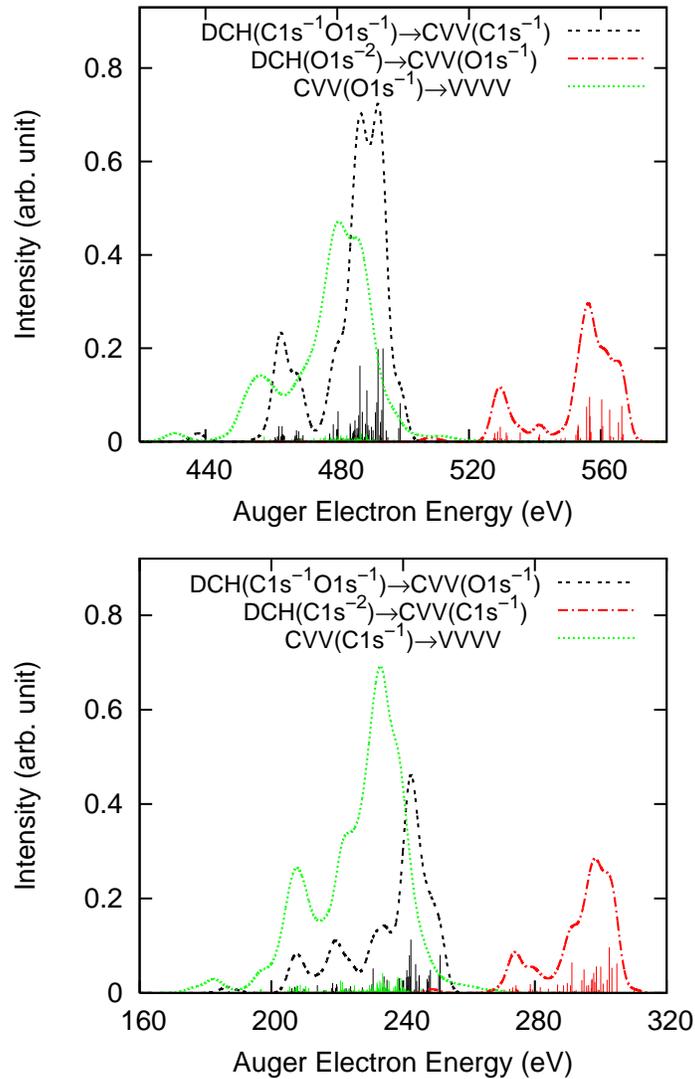}%
 \caption{\label{fig7}
 (Upper panel) The Auger intensities for the O1s core-hole decays of H$_2$CO molecule, 
 which include the Auger decays of the H$_2$CO O1s$^{-2}$ ss-DCH state, 
 the C1s$^{-1}$O1s$^{-1}$ ts-DCH states, and the O1s$^{-1}$ CVV states. 
 (Lower panel)  The Auger intensities for the C1s core-hole decays of H$_2$CO molecule, 
 which include the Auger decays of the H$_2$CO C1s$^{-2}$ ss-DCH state, 
 the C1s$^{-1}$O1s$^{-1}$ ts-DCH states, and the C1s$^{-1}$ CVV states. 
 The other details are the same as in Fig. \ref{fig1}. 
   }
\end{figure}

In Fig. \ref{fig7}, the Auger electron kinetic energy distribution 
of H$_2$CO is shown for the sequential Auger decays of the DCH and CVV states. 
In contrast to NH$_3$ and CH$_4$, H$_2$CO has four different 
DCH states: two ss-DCH states ( singlet C1s$^{-2}$ and O1s$^{-2}$ ) and two ts-DCH states 
( singlet and triplet C1s$^{-1}$O1s$^{-1}$ ). 
We consider the case where these four DCH states are created in XFEL experiment. 
In order to simplify the situation, we just assumed that the formation probabilities 
are determined by statistical ratio: 1:3 for singlet and triplet DCH states. 
This approximation has been well known and discussed in the literate of 
inner-shell photoionization.\cite{PhysRevA.9.1090,ISI:000231310500028} 
The formation probabilities of the singlet ts-DCH and the ss-DCH states 
are assumed to be the same, because they are expected to be formed through sequential two-photon 
two-electron ionization. 
We also calculated the Auger spectra expected in SR experiment, where only the ss-DCH states 
contribute the Auger spectra because of very low formation probabilities of the ts-DCH states. 
The result is shown in Fig. S3.  

The lower panel of Fig. \ref{fig7} shows the Auger spectra associated with the C1s 
core-hole decays, which include the Auger transition from the C1s$^{-2}$ ss-DCH state to 
the C1s$^{-1}$ CVV states, the transitions from the C1s$^{-1}$O1s$^{-1}$ ts-DCH states to 
the O1s$^{-1}$ CVV states, and the transitions from the C1s$^{-1}$ CVV states to 
the VVVV states. 
Similarly, the upper panel of Fig. \ref{fig7} shows the Auger spectra associated with 
the O1s core-hole decays, including the Auger transition from the O1s$^{-2}$ ss-DCH state to 
the O1s$^{-1}$ CVV states, the transitions from the C1s$^{-1}$O1s$^{-1}$ ts-DCH states 
to the C1s$^{-1}$ CVV states, and the transitions from the O1s$^{-1}$ CVV states to 
the VVVV states. 
Since the O1s$^{-1}$ CVV states are generated by the Auger decays of the O1s$^{-2}$ ss-DCH and 
the C1s$^{-1}$O1s$^{-1}$ ts-DCH states, the O1s$^{-1}$ CVV $\to$ VVVV Auger spectrum has two different 
origins. Likewise, the Auger decays of the C1s$^{-2}$ ss-DCH and the C1s$^{-1}$O1s$^{-1}$ ts-DCH states 
contribute to the C1s$^{-1}$ CVV $\to$ VVVV Auger spectrum. 
The DCH $\to$ CVV and CVV $\to$ VVVV Auger decays of H$_2$CO involve mainly 6 valence 
orbitals, i.e., three $a_1$, one $b_1$ and two $b_2$ MOs, because the main valence 
electron configuration 
of the DCH states is represented as $(3a_1)^2(4a_1)^2(1b_2)^2(5a_1)^2(1b_1)^2(2b_2)^2$.
Since this number of orbitals is larger than those in the CH$_4$ and NH$_3$ Auger decays, 
the discrete H$_2$CO Auger intensities in Fig. \ref{fig7} are more densely distributed  
in comparison with the intensities for the CH$_4$ and NH$_3$ Auger decays. 
Also, peaks in the convoluted H$_2$CO Auger spectra are not so distinct as in the spectra of 
the CH$_4$ and NH$_3$ Auger decays. 

As shown in the lower panel of Fig. \ref{fig7}, the main part of 
the C1s$^{-2}$ ss-DCH $\to$ C1s$^{-1}$ CVV Auger 
spectrum extends from 270 to 310 eV, 
with weak intensities around 250 eV. The C1s$^{-1}$O1s$^{-1}$ ts-DCH $\to$ O1s$^{-1}$ CVV Auger 
spectrum, including singlet and triplet contributions, extends 
from 210 to 250 eV, with weak intensities around 190 eV. 
The spectrum of the C1s$^{-1}$ CVV $\to$ VVVV Auger decays extends from 180 to 
260 eV. As can be seen in the figure, the Auger spectrum 
of the C1s$^{-2}$ ss-DCH $\to$ C1s$^{-1}$ CVV decay is well separated from the other spectra. 
However, the main parts of the C1s$^{-1}$O1s$^{-1}$ ts-DCH $\to$ O1s$^{-1}$ CVV and 
the C1s$^{-1}$ CVV $\to$ VVVV Auger spectra overlap each other. 
The main part of the O1s$^{-2}$ ss-DCH $\to$ O1s$^{-1}$ CVV Auger spectrum 
in the upper panel of Fig. \ref{fig7} 
extends from 510 to 570 eV, whereas the C1s$^{-1}$O1s$^{-1}$ ts-DCH $\to$ C1s$^{-1}$ CVV Auger 
spectrum extends from 460 to 500 eV. 
As in the case of the C1s$^{-1}$ core-hole decays, 
the spectrum of the O1s$^{-1}$ CVV $\to$ VVVV Auger decay overlaps 
with those of the C1s$^{-1}$O1s$^{-1}$ ts-DCH $\to$ C1s$^{-1}$ CVV Auger spectrum. 
Our results suggest that the Auger decays of the ss-DCH states 
may be easily identified in experiment. 
However, distinction between the ts-DCH $\to$ CVV Auger spectra and the CVV $\to$ VVVV 
Auger spectra may be difficult. 

The convoluted C1s$^{-2}$ ss-DCH $\to$ C1s$^{-1}$ CVV Auger spectrum has two large peaks at 
300 eV and 275 eV, and one faint peak at 250 eV. The peak at 300 eV is formed by two-electron 
vacancy creation in the 5$a_1$, 1$b_1$, 1$b_2$ and 2$b_2$ valence MOs, 
$(5a_{1} 1b_{1} 1b_{2} 2b_{2})^{-2}$, 
while the peak at 275 eV corresponds to the $(3a_1)^{-1}(5a_{1} 1b_{1} 1b_{2} 2b_{2})^{-1}$
valence hole creation. 
The structure between the 300 eV peak and the 275 eV peak involves vacancy 
creation in the 4$a_1$ MO as well. The faint peak at 250 eV is formed by 
the $(3a_1)^{-2}$ vacancy creation. Note that these assignments are not as obvious as 
in the cases of the NH$_3$ and CH$_4$ Auger spectra. 
Similarly, we can relate the peaks in the other convoluted Auger spectra to 
the $(5a_{1} 1b_{1} 1b_{2} 2b_{2})^{-2}$, 
$(3a_1)^{-1}(5a_{1} 1b_{1} 1b_{2} 2b_{2})^{-1}$ and $(3a_1)^{-2}$ vacancy creations,  
with contribution of the $4a_1$ hole creation to the structure between the highest energy and 
the second highest energy peaks. 
For example, the peak around 560 eV in the O1s$^{-2}$ ss-DCH $\to$ O1s$^{-1}$ CVV Auger 
spectrum and the peaks around 240 eV and 480-490 eV 
in the C1s$^{-1}$O1s$^{-1}$ ts-DCH $\to$ CVV Auger spectra correspond to 
the $(5a_{1} 1b_{1} 1b_{2} 2b_{2})^{-2}$ valence hole creation. 
Also, the peaks around 530-540 eV in the O1s$^{-2}$ ss-DCH $\to$ O1s$^{-1}$ CVV Auger 
spectrum and the peaks around 210-220 eV and 460-470 eV in the C1s$^{-1}$O1s$^{-1}$ 
ts-DCH $\to$ CVV Auger spectra are formed from the $(3a_1)^{-1}(5a_{1} 1b_{1} 1b_{2} 2b_{2})^{-1}$ 
vacancy creation. 
The origin of the peaks in the convoluted CVV $\to$ VVVV Auger spectra were not inspected 
because these peaks contain too many discrete transitions. Since the shapes 
of the convoluted CVV $\to$ VVVV spectra are roughly similar to the DCH $\to$ CVV Auger spectra, 
origin of the peaks may be the same as in the DCH $\to$ CVV Auger spectra. 

\begin{figure}
 \includegraphics[scale=1.5]{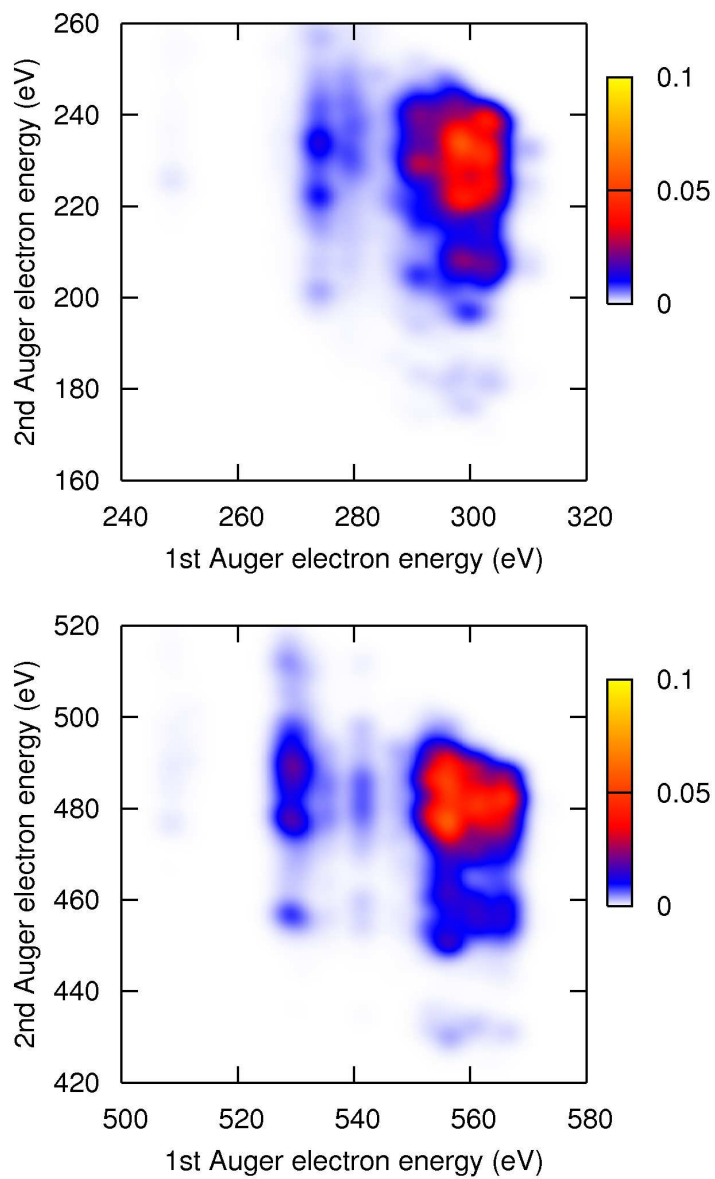}%
 \caption{\label{fig8} 
 (Upper panel) 2D intensity of the two-step Auger decays of the H$_2$CO C1s$^{-2}$ ss-DCH state, 
 as functions of kinetic energies of the 1st and the 2nd Auger electrons.
 (Lower panel) 2D intensity of the two-step Auger decays of the H$_2$CO O1s$^{-2}$ ss-DCH state, 
 as functions of kinetic energies of the 1st and the 2nd Auger electrons. 
 The other details are the same as in Fig. \ref{fig2}. 
   }
\end{figure}

\begin{figure}
 \includegraphics[scale=1.5]{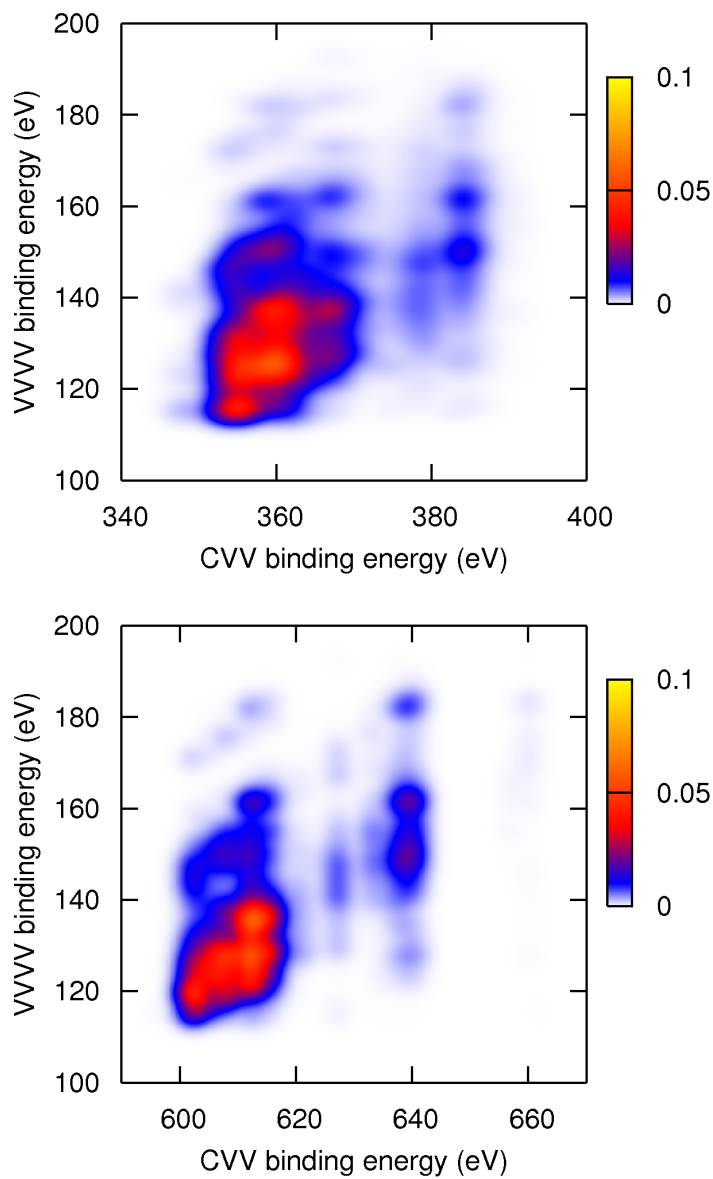}%
 \caption{\label{fig9} 
 (Upper panel) 2D Auger intensity of the H$_2$CO C1s$^{-2}$ ss-DCH decay 
 as functions of CVV and VVVV binding energies, converted from the 2D spectrum in the upper panel 
 of Fig. \ref{fig8}. 
 (Lower panel) 2D Auger intensity of the H$_2$CO O1s$^{-2}$ ss-DCH decay 
 as functions of CVV and VVVV binding energies, converted from the 2D spectrum in the lower panel 
 of Fig. \ref{fig8}. 
 The other details are the same as in Fig. \ref{fig2}. 
    }
\end{figure}

In the upper panel of Fig. \ref{fig8}, 2D Auger intensity distribution for the C1s$^{-2}$ 
ss-DCH $\to$ C1s$^{-1}$ CVV Auger decay and the subsequent C1s$^{-1}$ CVV $\to$ VVVV Auger 
decay is shown as functions of the 1st and the 2nd Auger electron kinetic energies.
Similarly, 2D Auger spectrum for the O1s$^{-2}$ ss-DCH $\to$ O1s$^{-1}$ CVV and the O1s$^{-1}$ 
CVV $\to$ VVVV Auger decays is shown in the lower panel of Fig. \ref{fig8}.
These 2D Auger spectra for the ss-DCH decays may be available in experiment 
using 4-electron coincidence technique, since the Auger spectra originated 
from the ss-DCH $\to$ CVV Auger decays are separated from the other Auger spectra   
as seen in Fig. \ref{fig7}. 
Compared to the peaks in the 2D spectra of the CH$_4$ and NH$_3$ DCH Auger decays, 
the peaks in the H$_2$CO ss-DCH decays look less clear. 
The highest intensity regions are located around the 1st Auger energies of 
295 - 305 eV and the 2nd Auger energies of 220 - 240 eV for the C1s$^{-2}$ case, and around 
the 1st Auger energies of 550 - 565 eV and the 2nd Auger energies of 470 - 490 eV for 
the O1s$^{-2}$ case. Location of these highest intensity regions indicates that 
the $(5a_{1} 1b_{1} 1b_{2} 2b_{2})^{-2}$ valence hole creation and subsequent 
$(5a_{1} 1b_{1} 1b_{2} 2b_{2})^{-2}$ valence hole creation 
are the most probable in the successive two Auger decays. 
The 2D Auger spectra in Fig. \ref{fig8} are converted to the 2D 
intensities as functions of CVV and VVVV binding energies as shown in Fig. \ref{fig9}. 
Based on these spectra, the highest-intensity peaks can be related to the low-lying 
CVV and VVVV states. In addition, we can recognize that some CVV states with moderately high 
energies, 385 eV in the C1s$^{-2}$ case and 640 eV in the O1s$^{-2}$ case, contribute 
to the higher-intensity peaks in Fig. \ref{fig9}. 
These 2D intensities as functions of CVV and VVVV binding energies may be 
available in experiment, if the binding energies of the ss-DCH states as well as the 2D 
Auger intensities as functions of the 1st and 2nd Auger electron energies are measured. 

\begin{figure}
 \includegraphics[scale=1.2]{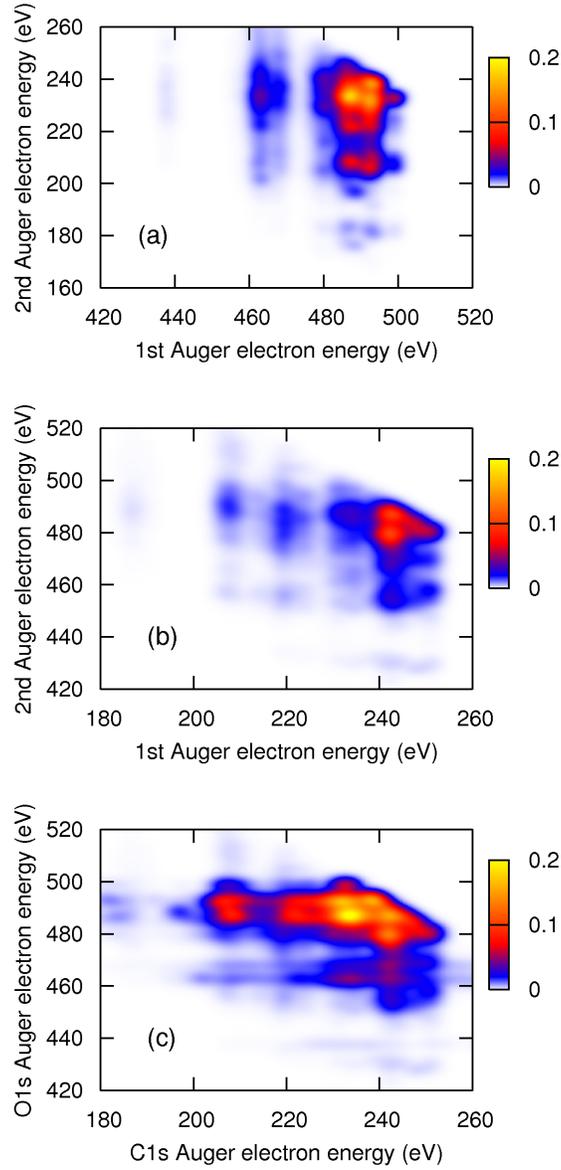}%
 \caption{\label{fig10} 
 2D Auger spectra of the two-step Auger decays of the H$_2$CO C1s$^{-1}$O1s$^{-1}$
 ts-DCH states. Panel (a): 2D spectrum with the 1st Auger transitions from 
 the C1s$^{-1}$O1s$^{-1}$ state to the C1s$^{-1}$ CVV states and the 2nd transitions from the C1s$^{-1}$ 
 CVV states to the VVVV states. Panel (b): 2D spectrum with the 1st Auger transitions from 
 the C1s$^{-1}$O1s$^{-1}$ state to the O1s$^{-1}$ CVV states and the 2nd transitions from the O1s$^{-1}$ 
 CVV states to the VVVV states. Panel (c): 2D Auger intensity as functions of C1s and 
 O1s Auger electron energies, obtained by adding the intensities of two different Auger 
 decay pathways in the panel (a) and (b).   
   }
\end{figure}

In the two-step Auger decays of the C1s$^{-1}$O1s$^{-1}$ ts-DCH states, 
two different Auger decay pathways exist: (1) the 1st Auger transitions 
from the ts-DCH states to the O1s$^{-1}$ CVV states, and subsequent Auger 
transitions from the O1s$^{-1}$ CVV states to the VVVV states, (2) 
the 1st Auger transitions from the ts-DCH states to the C1s$^{-1}$ CVV states, 
and subsequent Auger transitions from the C1s$^{-1}$ CVV states to the VVVV states. 
In the case (1), the 1st Auger electron has an energy in the region of the C1s core-hole decay 
and the 2nd Auger energy is in the region of the O1s core-hole decay. 
In contrast, in the case (2), the 1st Auger electron has an energy in the region of the O1s core-hole decay 
and the 2nd Auger energy is in the region of the C1s core-hole decay. 
The 2D Auger intensity distributions originated from these two different Auger decay pathways 
are shown in the panel (a) and (b) of Fig. \ref{fig10} as functions of the 1st and 2nd Auger electron 
energies. 
It may be difficult in experiment to determine the order of two successive Auger decays, or 
in other word, to make distinction between these two Auger decay pathways. 
Thus, two Auger intensities of the case (1) and (2) will not be observed separately, 
but summation of these two intensities will be measured as shown in the panel (c) of Fig. \ref{fig10}. 
The highest intensity region is located around the C1s Auger energy of 230 - 240 eV and the O1s 
Auger energy of 490 eV in the panel (c). 
As in the cases of the ss-DCH $\to$ CVV Auger decays, this region roughly corresponds to 
4-electron valence vacancy creation in the $5a_{1}$, $1b_{1}$, $1b_{2}$ and $2b_{2}$ orbitals 
in the sequential two Auger decays.  
Although other peaks and structures appear in the 2D intensity distribution, 
it is difficult to make clear assignment because two different Auger intensities are added 
in the panel (c). 
In contrast to the 2D Auger spectra of the ss-DCH decays, 
the 2D Auger spectrum of the ts-DCH decays in the panel (c) cannot be converted to the intensity 
as functions of CVV and VVVV binding energies, 
because such conversion requires individual 2D Auger spectra as in the panels (a) and (b). 
The integrated 1D Auger spectra as a function of VVVV binding energy, 
as in the lower panels of Figs. \ref{fig3} and \ref{fig6}, 
can be obtained from the 2D spectra in the panels (a) and (b), 
using the relation $E_{\rm VVVV} = E_{\rm DCH} - E_{\rm Auger1} - E_{\rm Auger2}$. 
Because this relation is preserved by exchange of $E_{\rm Auger1}$ and $E_{\rm Auger2}$, 
the 1D Auger spectrum can be directly obtained from the 2D spectrum in the panel (c). 
Thus, if experimental value of $E_{\rm DCH}$ is available, 
information on VVVV binding energies may be obtained in experimental 
measurement on the ts-DCH Auger decays. 

For reference, ionization energies of the low-lying C1s$^{-1}$ CVV states 
are shown in Table \ref{tab3} with their main configurations and Auger intensities. 
Also, the low-lying O1s$^{-1}$ CVV states are listed in Table \ref{tab4}. 
As in the ss-DCH $\to$ CVV Auger decays of NH$_3$ and CH$_4$, the singlet DCH states 
decay only to the doublet CVV states. In addition to these doublet states,  
the triplet C1s$^{-1}$O1s$^{-1}$ ts-DCH states decay to the quartet CVV states as well. 
In the DCH $\to$ CVV Auger decays of NH$_3$ and CH$_4$, the Auger intensity 
tends to be larger for the CVV final ion state with singlet intermediate valence spin 
than the state with triplet intermediate valence spin.  
However, this tendency is not obvious in H$_2$CO. 

\subsection{Discussion}

Our calculation on NH$_3$ DCH Auger spectrum has roughly reproduced 
the experimental Auger spectrum of Eland et al.,\cite{ISI:000284407400011} 
and thus similar agreements are expected for CH$_4$ and H$_2$CO DCH Auger spectra. 
The convoluted Auger spectra contain several large distinct peaks, 
which can be related to two hole creations in valence MOs. 
In case of the integrated 1D Auger spectra, we can relate their peaks to the 
electronic configurations of the CVV or VVVV states. 
Although this kind of assignment can be efficiently performed for small molecules 
such as CH$_4$ and NH$_3$, it may be difficult for larger molecule 
with many occupied valence MOs. 

Since there are many CVV and VVVV states in CH$_4$,NH$_3$ and H$_2$CO, 
possible number of transitions in the CVV $\to$ VVVV Auger decays might be huge, 
and thus distributions of the Auger spectra might be much broader and less structured 
compared to those of the DCH $\to$ CVV Auger decays. 
In contrast to this intuitive expectation, the distributions of the calculated CVV $\to$ VVVV 
Auger spectra have similar extent as in the DCH $\to$ CVV Auger spectra, 
as can be seen in Figs. \ref{fig1}, \ref{fig4} and \ref{fig7}. 
In addition, we can see several distinct peaks in the convoluted CVV $\to$ VVVV Auger 
spectra, which is especially pronounced in the CH$_4$ case. 
Our results can be understood in the following manner. 
When we evaluated the DCH $\to$ CVV Auger intensities using Eqs. (1) and (6), we approximated 
the DCH state by a single CSF function. This means that the DCH $\to$ CVV Auger intensity is 
noticeably large only when the CI wave function of the CVV state is dominated 
by a configuration obtained by removing two valence electrons from and adding 
one core electron to the DCH configuration, because only such a configuration 
has non-zero Auger amplitude in Eq. (1). 
Similarly, the CVV $\to$ VVVV Auger intensity is noticeably large only when 
(1) the CVV state has large intensity in the DCH $\to$ CVV Auger decay and 
(2) the CI wave function of the VVVV state is dominated by a configuration 
obtained by removing two valence electrons from and adding one core electron to the 
main configuraton of the CVV state. This also means that the final VVVV ion states are all 
dominated by electronic configurations constructed from the occupied valence MOs of the DCH state. 
The above conditions (1) and (2) severely restrict possible pairs of CVV and VVVV states 
which have large contribution to the CVV $\to$ VVVV Auger decay. 
The distribution of the CVV $\to$ VVVV Auger spectra cannot be as broad as 
intuitively expected from the distribution of the CVV and VVVV binding energies, 
because of the selection conditions stated above. 
These conditions also indicate that the Auger electron kinetic energy 
in the CVV $\to$ VVVV Auger decay is related to the two-electron valence hole creation 
in the occupied MOs of the DCH state. In other word, the CVV $\to$ VVVV Auger 
intensity can be noticeably large only around the specific energy region 
related to the two-electron valence hole creation, 
although this property may be less pronounced as number of occupied MOs increases. 

The total intensity of the H$_2$CO ts-DCH $\to$ C1s$^{-1}$ CVV Auger transitions,
obtained by integrating the spectrum by the Auger kinetic energy, 
is larger than that of the H$_2$CO ts-DCH $\to$ O1s$^{-1}$ CVV Auger transitions, as 
seen in Fig. \ref{fig7}. This difference indicates that the occupied valence MOs 
of the ts-DCH states have larger overlap with the oxygen site than that with the carbon site. 
If individual, separated experimental spectra are available for 
ts-DCH $\to$ CVV, ss-DCH $\to$ CVV and CVV $\to$ VVVV Auger transitions of H$_2$CO, 
and if assignment of peaks is available in terms of the occupied valence MOs, 
we can compare intensities of peaks having the same MO assignment in different Auger transitions. 
Such comparison may be useful in understanding relaxation of the valence MOs and overlap of MOs 
with different atomic sites. 
Similar analysis for ss- and ts-DCH Auger decays of other molecules will be interesting,  
if experimental measurement is possible. 


In this work, the Wentzel's ansatz was employed to estimate intensities 
of the 1st Auger transition from DCH to CVV states, and of the 2nd Auger 
transition from CVV to VVVV states. 
This means that we assume a ``three-step'' model in which 
DCH state creation by photoionization, the 1st Auger decay, and the 2nd Auger decay 
are all treated independently, 
as in the two-step model of single core-hole Auger decay.
\cite{ISI:A1992MG04500001,ISI:000172182900011,ISI:000182320200012,ISI:000261431200022}
This assumption is valid when the lifetimes of the intermediate states, 
DCH and CVV states in this case, are sufficiently long that the interaction between 
photoelectron and the 1st Auger electron as well as the interaction between the 1st Auger electron and 
the 2nd Auger electron are negligibly small. 
According to the multiconfiguration Dirac-Fock calculation by Chen on atomic DCH state,
\cite{ISI:A1991FV18400032} 
the Auger transition rates of DCH states for the low-Z elements are much larger than 
the Auger rates of SCH states; in the case of neon, the Auger rate per K hole of the DCH state 
is about 42 $\%$ larger than the rate of the SCH state. 
The Auger lifetimes of the CH$_4$, NH$_3$ and H$_2$CO DCH states 
may also be short, but such values are not known so far. 
Thus, it will be interesting as well as necessary to verify the validity 
of the ``three-step'' assumption in future,   
for example, by performing calculation or experimental measurement on lifetime 
of DCH and CVV states, or by measuring angular distribution of Auger electrons 
in the molecular-frame as in the SCH Auger process.\cite{ISI:000172182900011,ISI:000261431200022}

Following Mitani et al.,\cite{ISI:000181521100001} 
we approximated two-electron integrals in Eq. (\ref{ciamp}), 
which involve core-hole, valence and Auger continuum orbitals, 
by the L\"owdin population of the valence MO in the integral.  
When an Auger transition amplitude is dominated by a pair of 
CSFs, one in the initial state and the other in the final state, this approximation 
may be valid as demonstrated by Mitani et al.\cite{ISI:000181521100001} 
However, when more than one pair of CSFs equally contributes to an Auger amplitude,  
this approximation can be less accurate because it does not necessarily preserve 
proper relative phases of two-electron integrals in the Auger amplitude. 
In our calculation, Auger amplitudes were dominated by one pair of CSFs in most cases 
and thus this approximation is expected to be valid. 
Still, refinement of this approximation is desirable.  

%

In this paper, we have studied basic properties of molecular DCH Auger decay 
by estimating Auger spectra originated from the normal Auger decays of 
the molecular DCH states. 
Angular distribution of the Auger electron was not studied in this work, 
though the experimental measurement has already been performed for the Auger 
decay of the N$_2$ ss-DCH state.\cite{ISI:000281072100003} 
Angular distribution of the Auger electron may contain additional information on 
valence holes created in DCH or CVV Auger decays, 
and will be an interesting subject for theoretical study in future. 
In addition to the normal Auger decay of molecular DCH state, 
the Auger decays of doubly core excited resonance state and shake-up satellite state 
have been studied experimentally \cite{ISI:000284407400011,ISI:000287357100010} 
and may deserve further theoretical study as well. 

At this point of time, experimental Auger spectra are only available for 
the Auger decay of ss-DCH state.
\cite{ISI:000281072100004,ISI:000284407400011,ISI:000287357100010}
However, the XFEL experiment on CO molecule has been performed at LCLS and   
the Auger spectra of the ts-DCH decay are being analyzed.\cite{LF2011}  
Thus, we can expect further experimental information in this subject in 
the near future. 

\section{Summary}

We have performed theoretical investigation on Auger decay of 
molecular DCH state, for which experimental information 
becomes available for several molecules recently. 
Assuming sequential two-step Auger transitions from DCH state to CVV states and from 
CVV states to VVVV states, Auger electron kinetic energy distributions were 
estimated for CH$_4$, NH$_3$ and H$_2$CO molecules based on the CASSCF 
and CASCI calculations. 
The Auger spectra of the CH$_4$ and NH$_3$ ss-DCH Auger decays contain 
two well separated components: one from the 1st Auger transition from 
the DCH state to the CVV states and the other from the 2nd Auger transition 
from the CVV states to the VVVV final ion states. 
Our result roughly agrees with the experimental Auger spectra of NH$_3$ 
ss-DCH decay,\cite{ISI:000284407400011} but experimental spectrum with 
better energy resolution is desired for precise comparison. 
The calculated Auger spectrum of H$_2$CO DCH decay has more complicated structure 
compared to the spectra of CH$_4$ and NH$_3$ DCH decays, due to existence of 
the ts-DCH Auger decay. In the Auger spectrum of H$_2$CO DCH decay, the components 
originated from the ss-DCH $\to$ CVV Auger decays are well separated from 
the rest of the spectrum. However, the components originated from the 
ts-DCH $\to$ CVV Auger decays, and the components from the CVV $\to$ VVVV Auger decays 
overlap each other, making separation of the spectra difficult. 
The 2D Auger spectrum may be helpful in resolving this difficulty. 
We hope our calculation and analysis on the results, especially the H$_2$CO DCH Auger spectrum, 
will be useful in interpreting experimental result expected in the near future. 


\begin{acknowledgments}
We thank E.H.D. Eland for sending us the experimental Auger spectrum 
of the NH$_3$ DCH Auger decays. 
M.T. also thank N. Kosugi for discussion on theoretical method.  
M.E. acknowledge the support from JST-CREST and a Grant-in-Aid for
Scientific Research from the JSPS. 
\end{acknowledgments}

\clearpage


%

\clearpage

\listoffigures

\clearpage

\clearpage

\clearpage

\clearpage

\clearpage

\clearpage

\clearpage

\clearpage

\clearpage

\clearpage

\clearpage

\renewcommand{\thefigure}{S1}

\begin{figure}
 \includegraphics[scale=1.2]{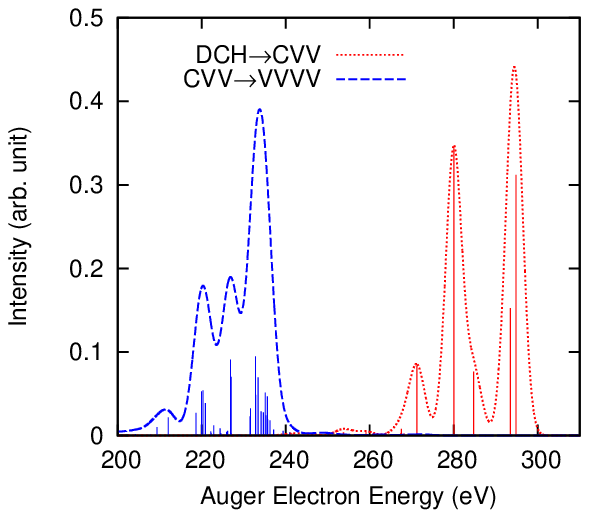}%
 \caption{\label{figS1} 
Effect of number of CI coefficients included in the calculation. 
We performed the same calculation as in Fig. 2, but with 30 configurations with the 
largest CI coefficients. 
The result is very similar to the CH$_4$ DCH Auger spectra in Fig. 2 which were 
calculated with 5 configurations with the largest CI coefficients. 
The details of the figure is the same as in the Fig. 2. 
   }
\end{figure}

\clearpage

\renewcommand{\thefigure}{S2}

\begin{figure}
 \includegraphics[scale=1.2]{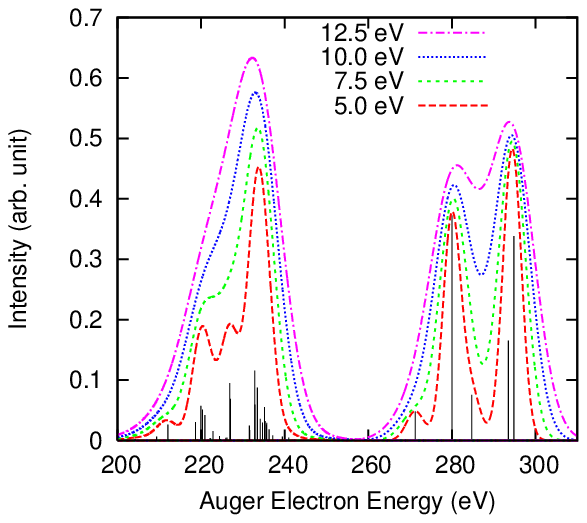}%
 \caption{\label{figS2} 
Effect of Gaussian width used for convolution.
The convoluted CH$_4$ DCH Auger spectra with different FWHM width are shown. 
Our result suggests that FWHM has strong influence on the shape of the Auger spectra, 
which may be important when we compare calculated spectrum with experimental result. 
The details of the figure is the same as in Fig. 2.
   }
\end{figure}

\clearpage

\renewcommand{\thefigure}{S3}

\begin{figure}
 \includegraphics[scale=1.2]{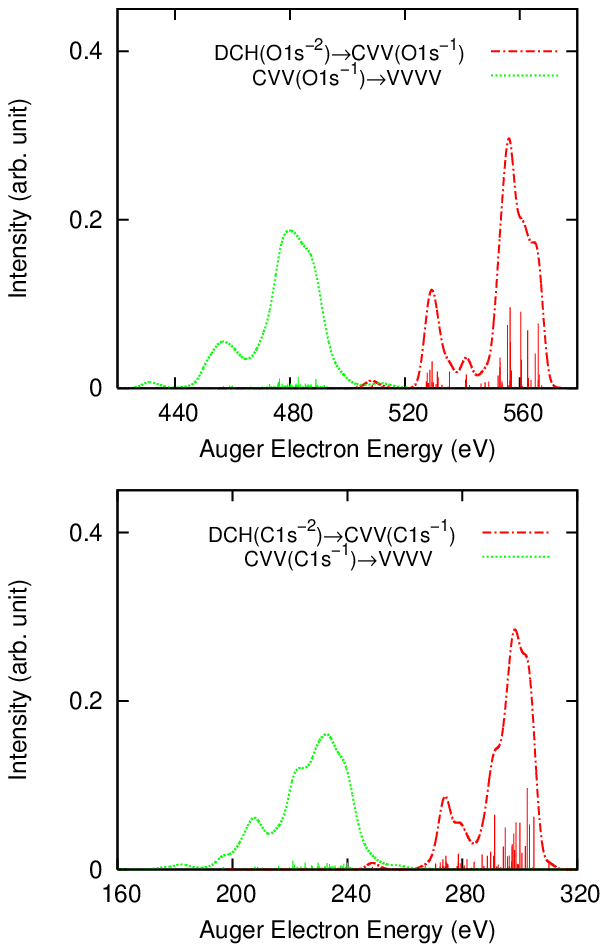}%
 \caption{\label{figS3}
H$_2$CO DCH $\to$ CVV and CVV $\to$ VVVV Auger spectra expected in SR experiment.
In Fig. 8, we show H$_2$CO DCH Auger spectra expected for XFEL experiment. 
In case of SR experiment, formation efficiencies of the ts-DCH states are expected to 
be low, 1/100 of the ss-DCH states. Thus, we can ignore the ts-DCH contributions in 
simulating Auger spectra in SR experiment. In this figure, the Auger spectra without 
the ts-DCH contribution is shown. The details of the figure is the same as in Fig. 8.
   }
\end{figure}

\clearpage

%


\clearpage

\begin{turnpage}

\begin{table}%
\caption{\label{tab0}
Expressions of Auger decay amplitude $t$ for transitions from ss-DCH, singlet ts-DCH and triplet ts-DCH states 
to CVV state, with assumptions of single configurational wave functions and frozen orbitals. 
``CVV S=1/2 (S)'' represents doublet CVV state with singlet intermediate spin coupling of valence 
electrons, and ``CVV S=1/2 (T)'' represents doublet CVV state with triplet intermediate spin coupling of valence 
electrons. $v$ and $w$ refer to orbitals involved in the valence hole creation, $c$ stands for inner-shell orbital 
involved in the core-hole decay, and $k$ represents continuum orbital of Auger electron.   
}
\begin{ruledtabular}
\begin{tabular}{l|ccc}
                       &  ss-DCH      & ts-DCH (singlet)  & ts-DCH (triplet)  \\
\hline
CVV S=1/2 (S)  & \shortstack{$\left( kv|cw \right) + \left( kw|cv \right) ~ \left( v \neq w \right)$ \\ $ \sqrt{2} \left( kv|cw \right) ~ \left( v = w \right)$}   & 
\shortstack{$\sqrt{\frac{1}{2}} \left[ \left( kv|cw \right) + \left( kw|cv \right) \right] ~ \left( v \neq w \right)$ \\ $\left( kv|cw \right) ~ \left( v = w \right)$}    &
\shortstack{$\sqrt{\frac{1}{2}} \left[ \left( kv|cw \right) + \left( kw|cv \right) \right] ~ \left( v \neq w \right)$ \\ $\left( kv|cw \right) ~ \left( v = w \right)$}    \\
CVV S=1/2 (T)  & $\sqrt{3} \left[ \left( kv|cw \right) - \left( kw|cv \right) \right]$  & $\sqrt{\frac{3}{2}} \left[ \left( kv|cw \right) - \left( kw|cv \right) \right]$  &    $\sqrt{\frac{1}{6}} \left[ \left( kv|cw \right) - \left( kw|cv \right) \right]$    \\
CVV S=3/2                  & $-$     &  $-$ &  $\sqrt{\frac{4}{3}} \left[ \left( kv|cw \right) - \left( kw|cv \right) \right]$      \\
\hline
\end{tabular}
\end{ruledtabular}
\end{table}

\end{turnpage}

\clearpage

\begin{table}%
\caption{\label{tab1}
 Representative CVV states of CH$_4$ molecule. Energy refers to the ionization energy 
 with respect to the neutral ground state of CH$_4$. 
 Intermediate spin means spin coupling in valence electrons, 
 where S and T represent singlet and triplet, respectively. 
 }
\begin{ruledtabular}
\begin{tabular}{cccc}
State & Energy (eV) & Main configuration (Intermediate spin) & Intensity (arb.) \\
\hline
 ${}^4T_1$   & 352.662 & $(1s)^1(1t_2)^4$         ~(T)        &     -     \\
 ${}^2E$     & 354.362 & $(1s)^1(1t_2)^4$         ~(S)        &    0.000  \\
 ${}^2T_1$   & 354.809 & $(1s)^1(1t_2)^4$         ~(T)        &    0.000  \\
 ${}^2T_2$   & 355.358 & $(1s)^1(1t_2)^4$         ~(S)        &    0.338  \\
 ${}^2A_1$   & 356.676 & $(1s)^1(1t_2)^4$         ~(S)        &    0.165  \\
 ${}^4T_2$   & 361.996 & $(1s)^1(2a_1)^1(1t_2)^5$ ~(T)        &     -     \\
 ${}^2T_2$   & 365.469 & $(1s)^1(2a_1)^1(1t_2)^5$ ~(T)        &    0.075  \\
 ${}^2T_2$   & 370.181 & $(1s)^1(2a_1)^1(1t_2)^5$ ~(S)        &    0.372  \\
 ${}^2A_1$   & 378.961 & $(1s)^1(2a_1)^0$         ~(S)        &    0.048  \\
\hline
\end{tabular}
\end{ruledtabular}
\end{table}

\clearpage

\begin{table}%
\caption{\label{tab2}
 Representative CVV states of NH$_3$ molecule. 
 The other details are the same as in Table \ref{tab1}.  
 }
\begin{ruledtabular}
\begin{tabular}{cccc}
State & Energy (eV) & Main configuration (Intermediate spin)  & Intensity (arb.)\\
\hline
 ${}^4E$     & 470.431 &  $(1s)^1(1e)^3(3a_1)^1$ ~(T)          &  -       \\
 ${}^2A_{1}$ & 470.810 &  $(1s)^1(3a_1)^0$       ~(S)          &  0.078   \\
 ${}^2E$     & 474.026 &  $(1s)^1(1e)^3(3a_1)^1$ ~(S)          &  0.168   \\
 ${}^2E$     & 475.207 &  $(1s)^1(1e)^3(3a_1)^1$ ~(T)          &  0.154   \\
 ${}^4A_{2}$ & 476.080 &  $(1s)^1(1e)^2$         ~(T)          &  -       \\
 ${}^2A_{2}$ & 479.240 &  $(1s)^1(1e)^2$         ~(T)          &  0.000   \\
 ${}^2E$     & 479.320 &  $(1s)^1(1e)^2$         ~(S)          &  0.134   \\
 ${}^2A_{1}$ & 480.593 &  $(1s)^1(1e)^2$         ~(S)          &  0.140   \\
 ${}^4A_{1}$ & 482.519 &  $(1s)^1(2a_1)^1(3a_1)^1$ ~(T)          &  -       \\
 ${}^4E$     & 487.973 &  $(1s)^1(2a_1)^1(1e)^1$   ~(T)          &  -       \\
 ${}^2A_{1}$ & 488.317 &  $(1s)^1(2a_1)^1(3a_1)^1$ ~(T)          &  0.049   \\
 ${}^2A_{1}$ & 491.578 &  $(1s)^1(2a_1)^1(3a_1)^1$ ~(S)          &  0.066   \\
 ${}^2E$     & 492.381 &  $(1s)^1(2a_1)^1(1e)^1$   ~(T)          &  0.056   \\
 ${}^2E$     & 497.398 &  $(1s)^1(2a_1)^1(1e)^1$   ~(S)          &  0.128   \\
 ${}^2A_{1}$ & 508.638 &  $(1s)^1(2a_1)^0$         ~(S)          &  0.020   \\
\hline
\end{tabular}
\end{ruledtabular}
\end{table}

\clearpage

\begin{turnpage}

\begin{table}%
\caption{\label{tab3}
 Low-lying C1s$^{-1}$ CVV states of H$_2$CO molecule. 
 C1s$^{-2}$, C1s$^{-1}$O1s$^{-1}$ (S) and C1s$^{-1}$O1s$^{-1}$ (T) are the initial states 
 of the 1st Auger transitions, and represent 
 the C1s$^{-2}$ ss-DCH state, the singlet 
 and triplet C1s$^{-1}$O1s$^{-1}$ ts-DCH states, respectively. 
 The other details are the same as in Table \ref{tab1}. 
 }
\begin{ruledtabular}
\begin{tabular}{ccc|ccc}
\multicolumn{3}{c|}{} & \multicolumn{3}{c}{Intensity (arb.)} \\
State & Energy (eV) & Main configuration (Intermediate spin) & C1s$^{-2}$ & C1s$^{-1}$O1s$^{-1}$ (S) & C1s$^{-1}$O1s$^{-1}$ (T) \\
\hline
 ${}^2A_1$ & 347.931 & $(2a_1)^{1}(2b_2)^{0}$           ~(S)  & 0.007 & 0.029 & 0.083 \\
 ${}^4A_2$ & 351.299 & $(2a_1)^{1}(1b_1)^{1}(2b_2)^{1}$ ~(T)  & -     & -     & 0.000 \\
 ${}^2A_2$ & 351.646 & $(2a_1)^{1}(1b_1)^{1}(2b_2)^{1}$ ~(T)  & 0.000 & 0.002 & 0.010 \\
 ${}^4A_1$ & 352.141 & $(2a_1)^{1}(1b_2)^{1}(2b_2)^{1}$ ~(T)  & -     & -     & 0.024 \\
 ${}^4B_2$ & 352.408 & $(2a_1)^{1}(2b_2)^{1}(5a_1)^{1}$ ~(T)  & -     & -     & 0.001 \\
 ${}^2B_2$ & 352.537 & $(2a_1)^{1}(2b_2)^{1}(5a_1)^{1}$ ~(T)  & 0.000 & 0.001 & 0.000 \\
 ${}^2A_1$ & 353.532 & $(2a_1)^{1}(1b_2)^{1}(2b_2)^{1}$ ~(T)  & 0.002 & 0.042 & 0.009 \\
 ${}^2A_2$ & 353.091 & $(2a_1)^{1}(1b_1)^{1}(2b_2)^{1}$ ~(S)  & 0.063 & 0.068 & 0.200 \\
 ${}^2B_2$ & 354.603 & $(2a_1)^{1}(2b_2)^{1}(5a_1)^{1}$ ~(S)  & 0.053 & 0.084 & 0.198 \\
 ${}^2A_1$ & 355.407 & $(2a_1)^{1}(1b_2)^{1}(2b_2)^{1}$ ~(S)  & 0.097 & 0.005 & 0.063 \\
\hline
\end{tabular}
\end{ruledtabular}
\end{table}

\end{turnpage}

\clearpage

\begin{turnpage}

\begin{table}%
\caption{\label{tab4}
 Low-lying O1s$^{-1}$ CVV states of H$_2$CO molecule. 
 The other details are the same as in Table \ref{tab3}. 
 }
\begin{ruledtabular}
\begin{tabular}{ccc|ccc}
\multicolumn{3}{c|}{} & \multicolumn{3}{c}{Intensity (arb.)} \\
State           & Energy (eV) & Main configuration (Intermediate spin) & O1s$^{-2}$ & C1s$^{-1}$O1s$^{-1}$ (S) & C1s$^{-1}$O1s$^{-1}$ (T) \\
\hline
 ${}^2A_1$ & 595.785 & $(1a_1)^{1}(2b_2)^{0}$           ~(S)  & 0.001 & 0.040 & 0.080 \\
 ${}^4A_2$ & 598.722 & $(1a_1)^{1}(1b_1)^{1}(2b_2)^{1}$ ~(T)  & -     & -     & 0.049 \\
 ${}^4A_1$ & 599.230 & $(1a_1)^{1}(1b_2)^{1}(2b_2)^{1}$ ~(T)  & -     & -     & 0.037 \\
 ${}^2A_2$ & 599.395 & $(1a_1)^{1}(1b_1)^{1}(2b_2)^{1}$ ~(S)  & 0.000 & 0.030 & 0.024 \\
 ${}^4B_2$ & 599.876 & $(1a_1)^{1}(2b_2)^{1}(5a_1)^{1}$ ~(T)  & -     & -     & 0.006 \\
 ${}^2B_2$ & 600.951 & $(1a_1)^{1}(2b_2)^{1}(5a_1)^{1}$ ~(T)  & 0.004 & 0.001 & 0.009 \\
 ${}^2A_1$ & 601.788 & $(1a_1)^{1}(1b_2)^{1}(2b_2)^{1}$ ~(T)  & 0.016 & 0.013 & 0.001 \\
 ${}^2B_2$ & 602.113 & $(1a_1)^{1}(2b_2)^{1}(5a_1)^{1}$ ~(S)  & 0.010 & 0.004 & 0.037 \\
 ${}^2A_2$ & 602.145 & $(1a_1)^{1}(1b_1)^{1}(2b_2)^{1}$ ~(T)  & 0.077 & 0.001 & 0.034 \\
 ${}^4B_2$ & 602.876 & $(1a_1)^{1}(2b_2)^{1}(4a_1)^{1}$ ~(T)  & -     & -     & 0.025 \\
 ${}^2A_1$ & 603.166 & $(1a_1)^{1}(1b_2)^{1}(2b_2)^{1}$ ~(S)  & 0.041 & 0.007 & 0.061 \\
\hline
\end{tabular}
\end{ruledtabular}
\end{table}

\end{turnpage}

\clearpage


\end{document}